\documentclass[aps,pra,floatfix,superscriptaddress,twocolumn]{revtex4}

\usepackage{graphicx,psfrag,color,bbm}
\usepackage[dvipsnames]{xcolor}
\usepackage{amsmath,amssymb}
\usepackage{mathtools}
\usepackage{hyperref}
\usepackage{bm}

\newcommand{\ket}[1]{|#1 \rangle}
\newcommand{\bra}[1]{\langle  #1 | }

\newcommand{\EAj}{E_A^{(j)}}

\newcommand{\average}[1]{\langle {#1} \rangle}

\newcommand{\tr}{\mathrm{tr}}

\newcommand{\Pigp}{\Pi^{(0)}_p}
\newcommand{\psigp}{E^{(0)}_{p}}

\definecolor{darkbrown}{rgb}{0.4, 0.26, 0.13}
\definecolor{darkmagenta}{rgb}{0.55, 0.0, 0.55}

\newcommand{\pg}[1]{\left\{{#1}\right\}}

\newcommand{\nickAdd}[1]{#1}
\newcommand{\alb}[1]{#1}
\newcommand{\albb}[1]{#1}
\newcommand{\gab}[1]{#1}

\usepackage{comment}

\begin{document}

\title{A thermodynamic approach to optimization in complex quantum systems}

\author{Alberto Imparato}
\affiliation{Department of Physics and Astronomy, Aarhus University, Denmark}
\author{Nicholas Chancellor}
\affiliation{Department of Physics and Joint Quantum Centre Durham-Newcastle, Durham University and Newcastle
University School of Computing, Urban Science Building NE4 5TG Newcastle, United Kingdom}
\author{Gabriele De Chiara}
\affiliation{Centre for Quantum Materials and Technology, School of Mathematics and Physics, Queen’s University Belfast, Belfast BT7 1NN, United Kingdom}

\begin{abstract}
We consider the problem of finding the energy minimum of a complex quantum Hamiltonian by employing a non-Markovian bath prepared in a low energy state.
The energy minimization problem is thus turned into a thermodynamic cooling protocol in which we repeatedly put the system of interest in contact with a {\it colder} auxiliary system.
By tuning the internal parameters of the bath, we show that the optimal cooling is obtained in a regime where the bath exhibits a quantum phase transition in the thermodynamic limit. This result highlights the importance of collective effects in thermodynamic devices. 
We furthermore introduce a two-step protocol that combines the interaction with the bath with a measure of its energy. While this protocol does not destroy coherence in the system of interest, we show that it can further enhance the cooling effect.
\end{abstract}
\maketitle

\section{Introduction} 
Optimization problems arise in different fields as computer science, physics, computational biology or drug design.
One possible approach to solve an optimization problem with quantum algorithms is to encode the problem into an Ising Hamiltonian
\begin{equation}
H_p=-\frac 1 2 \sum_{i,j} J_{ij} \sigma_i^z\sigma_j^z -\sum_i h_i \sigma_i^z,
\label{Hp:def}
\end{equation} 
where $\sigma_i^w$ is the Pauli $w=x,y,z$ operator of the $i$th qubit, while the coupling strength $J_{ij}$ and external field $h_i$ specify the optimization problem.
Models like the one in Eq.~\eqref{Hp:def} were originally introduced in condensed matter physics to study disordered magnetic materials \cite{Mezard1987}, but are now used to describe, e.g., neural networks \cite{Amit85}, protein folding \cite{Bryngelson87}, socioeconomic behavior such
as  drug and tobacco  use \cite{Durlauf1999} and trading models in stock markets \cite{Maskawa2002}. For reviews and perspectives on quantum computing applied to combinatorial optimisation, see \cite{yarkoni2022quantum,Au-Yeung2023,Abbas2023a}.

When the parameters $J_{ij}$ and $h_i$ in Eq.~\eqref{Hp:def} are normally distributed, the model in \eqref{Hp:def} becomes a Sherrington-Kirkpatrick (SK)
 spin-glass \cite{SherringtonPRL1975,Mezard1987}, whose energy minimization is an NP--hard problem.
NP--hard problems are the hardest class of classical optimisation problems  \cite{Garey1990,Papadimitriou1982}, examples include variants of the Traveling Salesman problem  \cite{Martin2001}. One disadvantage of NP--hardness as a concept is that it is  a worst case statement, proven by mapping other NP--hard problems using polynomial resources. As a result, NP--hardness actually tells us nothing about the difficulty of a ``typical'', in other words randomly generated, instance of a problem, and examples are known of NP--hard problems where typical instances are easy \cite{Beier2004a,Krivelevich2006a}. For numerical studies such as this one, we also want the additional property of {\it uniform} hardness, which is a statement about the hardness of typical instances. The presence of a finite-temperature spin-glass transition \cite{SherringtonPRL1975}  strongly suggests that as well as being NP-hard, the SK spin glass is uniformly hard, for this reason it has been used in other similar numerical studies \cite{Callison2019, Farhi2022}.

The use of quantum algorithms to solve such problems has the potential to provide a speedup over classical algorithms, provided that the system is made genuinely quantum by including a non-commuting driver term to the system Hamiltonian \cite{Choi2010,Lucas2014}. The difficulty in proving speedups for NP--hard problems derives from the fact that the best classical algorithms are not known (strictly speaking it is not even known if the scaling is exponential for the best algorithm, although it is strongly suspected to be). In the case of unstructured search a more artificial problem where the best algorithm is known, Hamiltonians based adiabatic \cite{Roland2002} and quantum walk \cite{childs2004spatial} approaches (and in fact a family of algorithms which interpolate between the two\cite{morley2017quantum}) are known to yield an optimal speedup equivalent to the one given by Grover's well known gate model algorithm \cite{Grover1996search,Grover1997search}.

\begin{figure}[t]
\center
\includegraphics[width=\columnwidth]{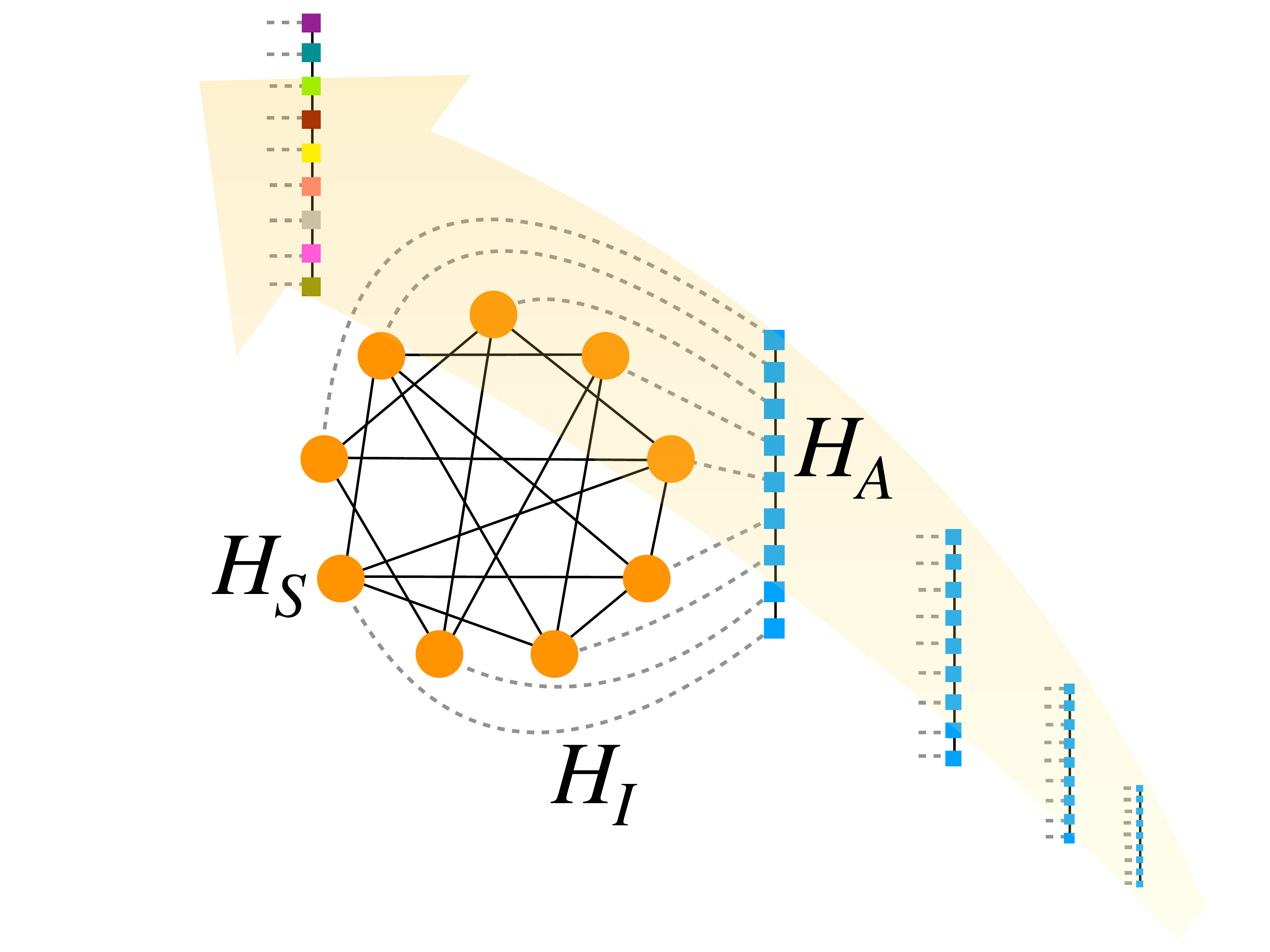}
\caption{Schematic of the collision protocol: a series of auxiliary spin chains (squares) interact with the SK spin glass system (circles). Black and red solid lines connecting spins in the glass represent positive and negative pairwise interactions, respectively, as described by the terms proportional to $J_{ij}$ in Eq.~\eqref{Hp:def}. Their thickness represents their random relative magnitude. The gray broken lines represent the interaction between the spins in the system and in the bath, as  given by $J$ in Eq.~\eqref{HI:def}. At the end of every collisional stroke the {\it exhaust} auxiliary chain is wasted and replaced by a new chain.} 
\label{fig:setup}
\end{figure}

In order to find the ground state of the model described by Eq.~\eqref{Hp:def}, in this paper we use an approach that has been termed {\it computing by cooling} in \cite{Feng2022}: in practice one employs a non-Markovian quantum bath in the form of a spin system with an easy-to-prepare low-temperature state, see a sketch of our proposal in Fig.~\ref{fig:setup}. 
Coupling the system represented by the problem Hamiltonian \eqref{Hp:def} to such a cold bath, the system of interest is thus cooled down toward its ground state. \gab{Other thermodynamic-inspired approaches to classical and quantum computation have been previously proposed~\cite{aifer2023thermodynamic, lipkabartosik2023thermodynamic}.}


As a novel approach in the present paper we consider baths that can undergo quantum phase transitions (QPT) in the thermodynamic limit. We show that by tuning the parameter that determines the bath phase, we can speed up the search for the ground state of the problem Hamiltonian. 
Second--order QPT and other collective phenomena have indeed been shown to
boost dynamic and thermodynamic performance in a number of thermal devices, both in classical and quantum regime \cite{Golubeva2012a, Imparato15, Campisi2016, Fusco2016, Ma2017, Herpich2018, Herpich2018a, Sune19a, Zhang2019, Chen2019, Abiuso2020, Imparato21, Puebla2022}.

While open system effects such as those experienced in our cooling approach could ruin the potential for quantum advantage, it is nevertheless known that even a projective measurement, which leads to pure decoherence, can lead to an optimal quantum speedup for unstructured search \cite{ChildsPRA2002} (the problem encountered by Grover's algorithm) if performed in the correct basis. While unstructured search has the advantage of being a setting where quantum computing has  known, provable speedup over classical computing, it has the disadvantage of behaving very different from the real optimisation problem considered in Ref.~\cite{Callison2019},  providing less information about possible performance on moderately sized real problems. For this reason, in this work, we elect to study a uniformly hard problem such as finding the minimum energy of SK spin glasses. \nickAdd{The reason uniform hardness is important is that it implies that random instances will be hard. This is not true for typical NP-hardness, since the statement of NP-hardness is based on the ability to map to other problems and is thus a worst-case rather than a typical-case statement. There are in fact well known instances of NP-hard problems where randomly generated instances are easy to solve \cite{Beier2004a,Krivelevich2006a}. Ising spin glasses however have a finite temperature spin glass transition, which strongly suggests that even random instances will be hard \cite{kirkpatrick1975solvable} (this property is known as uniform hardness). This subtlety had led to problems in previous works benchmarking quantum annealers, where it was realised that although solving Ising models when limited to the hardware graph of flux qubit devices is NP-hard, it is in fact not uniformly hard, and such problems are easy for classical Monte-Carlo algorithms \cite{katzgraber2014glassy}.}

This work is organised as follows. In Sec.~\ref{sec:pre} we start by describing our model while in Sec.~\ref{sec:computational} we discuss in detail the computational cooling scheme we propose. In Sec.~\ref{sec:measuring} we introduce a scheme which involves measuring the auxiliary system after its collision with the system of interest and in Sec.~\ref{sec:conclusions} we summarise and discuss our findings.

\section{Preliminaries}
\label{sec:pre}

In this paper we use the model described by the Hamiltonian in Eq.~\eqref{Hp:def} with random coefficients as a test-bed 
 to assess the performance of our algorithms and compare them to other quantum algorithms. \nickAdd{In particular, we study a modified Sherrington-Kirkpatrick spin glass, as referenced in the introduction, consisting of randomly coupling all spins with coupling strengths and fields in Eq.~\eqref{Hp:def} selected from a Gaussian distribution\footnote{The original SK model does not include fields, but they are convenient to remove degeneracy; it has been argued in \cite{Callison2019} that adding fields in this way does not have any effect on the hardness properties of the model.}.}
We start our analysis  by following the procedure described in Ref.~\cite{Chancellor2021} as a basis for comparison. 
\alb{The parameters $J_{ij}$ and $h_i$ are drawn from a normal distribution with zero mean and variance equal to one.}
In particular, we complement the problem Hamiltonian \eqref{Hp:def} with a ``driver'' Hamiltonian, and following previous works \cite{Choi2010,Lucas2014,Callison2019,Chancellor2021}, we will take it of the form 
\begin{equation}
H_d= -\sum_{i=1}^{N} \sigma_i^x,\\
\label{Hd:def}
\end{equation} 
such that the total Hamiltonian reads
\begin{equation}
H(\gamma(t))=\gamma(t) H_d + H_p.
\label{eq:Ht}
\end{equation} 
Such a Hamiltonian results in quantum tunneling among the
localized classical states, which correspond to the eigenstates $\ket{E^{(i)}_p}$ of $H_p$ (forming the computational basis).  

The ground state of $H_d$ is  the equal superposition of all the states forming the computational basis, a state that is in principle easy to prepare.
Indeed, at $t=0$, we prepare the system in the ground state of the driver Hamiltonian \eqref{Hd:def} and consider the following unitary dynamics generated by the Hamiltonian~\eqref{eq:Ht}, \gab{a process sometimes called a ``two-stage quantum walk"}. As time protocol we choose  $\gamma(t)=\gamma_1 +(\gamma_2-\gamma_1) \theta(t-t_q)$  i.e. a quench in  the driver's strength $\gamma$ at time $t_q$, where we have used the step function $\theta(t)$, which is 0 for $t<0$ and 1 for $t \ge 0$. The post-quench unitary dynamics of the total  system with $H(\gamma_2)$ corresponds thus to a quantum  random walk driven by $H_d$. 

\gab{Notice that the system, initially in the ground state of $H_d$, effectively experiences a quench also at $t=0$ when its Hamiltonian changes from $H_d$ to $H(\gamma_1)$. We emphasise the importance of the two quenches, one at $t=0$ and one at $t=t_q$, as means to change the system energy, which is otherwise conserved during the quantum walk dynamics.} 

We finally evaluate the expectation values of the different energy terms together with the fidelity with respect to the state of interest, as given by
\begin{equation}
P(t)= \tr \pg{\rho(t) \Pigp},
\label{fid:def}
\end{equation} 
with $\Pigp=\ket{\psigp} \bra{\psigp}$ the projector onto the problem Hamiltonian ground state.

\begin{figure}[h]
\center
\psfrag{ }[ct][ct][1.]{ }
\includegraphics[width=8cm]{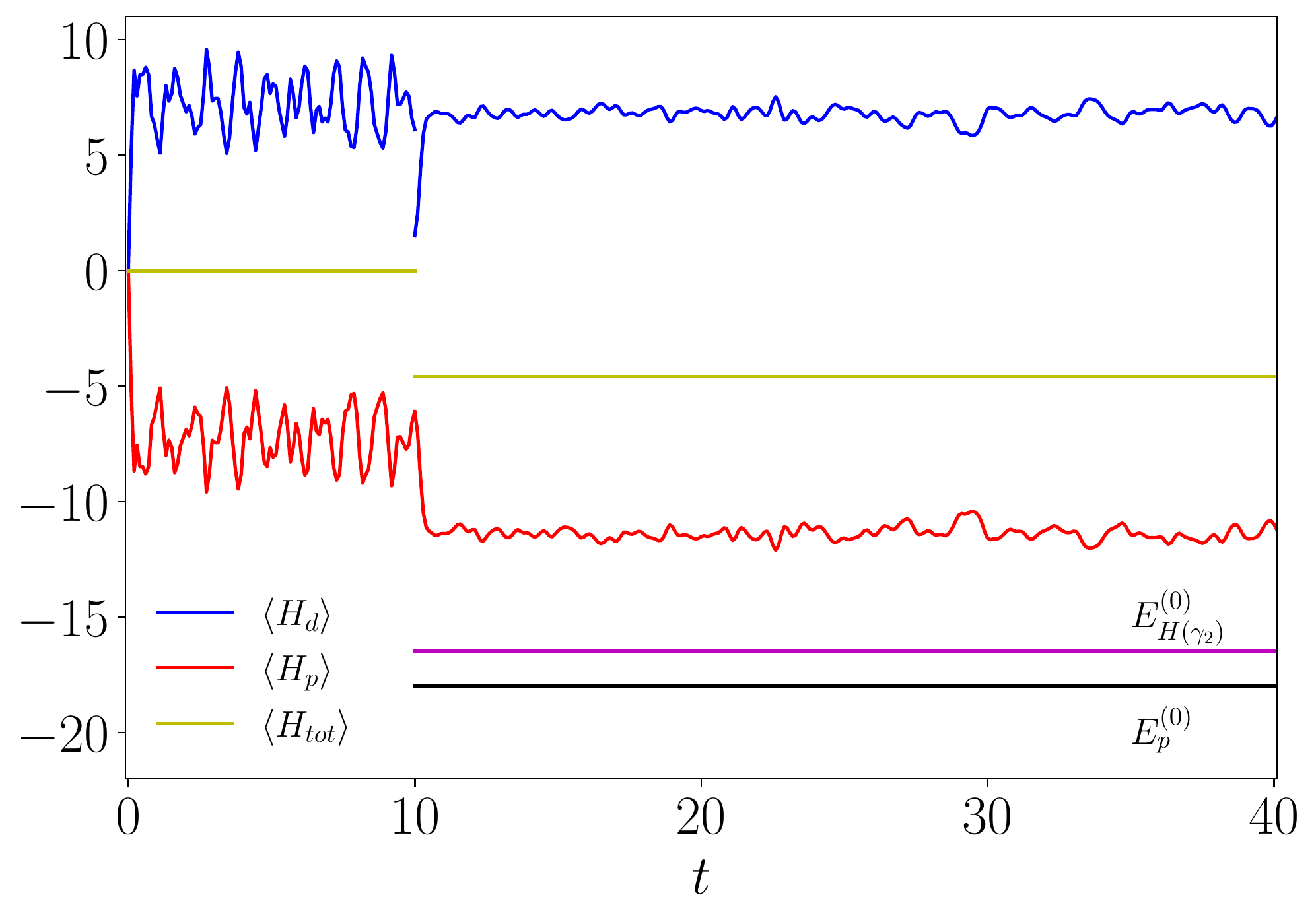}
\includegraphics[width=8cm]{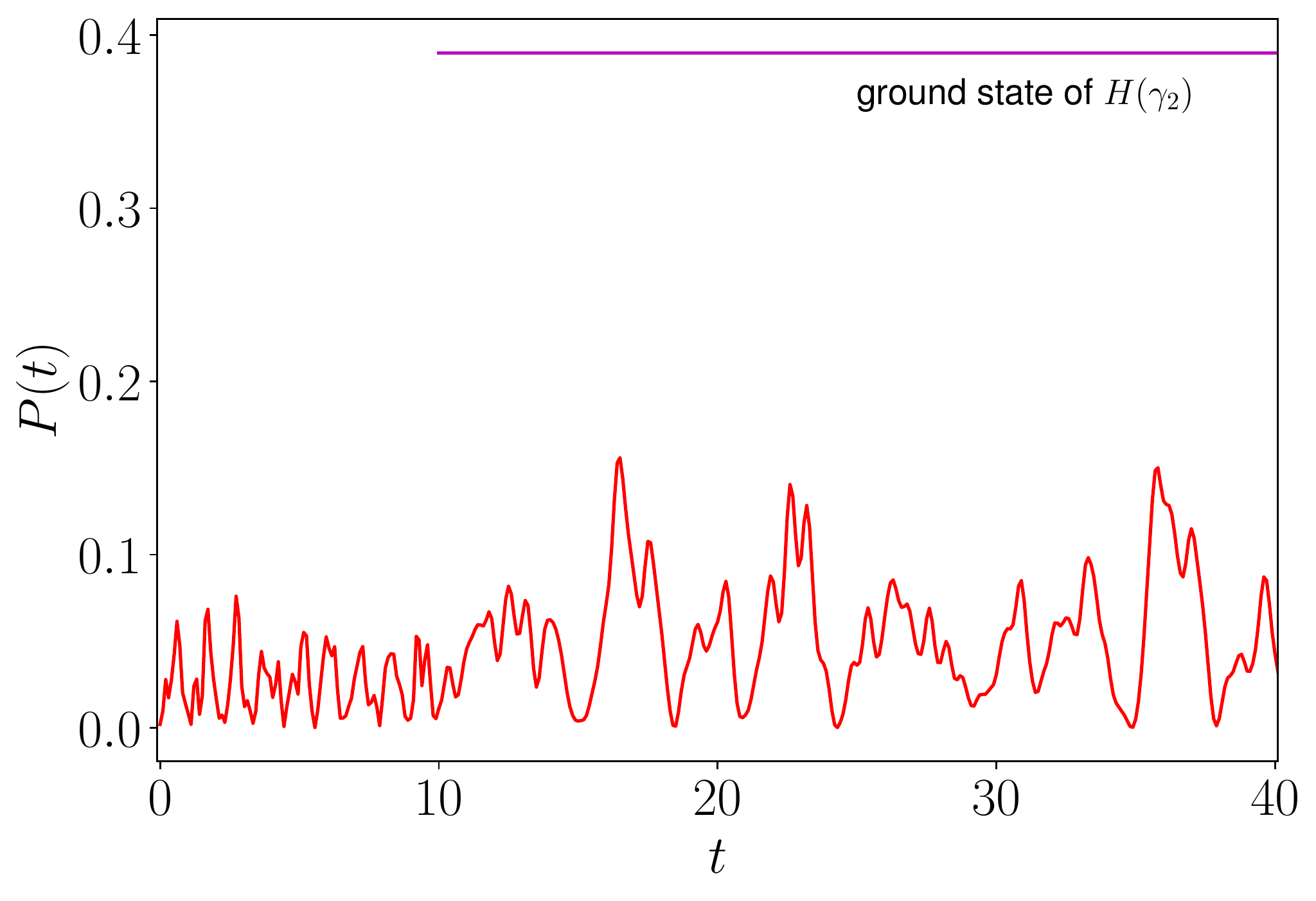}
\caption{Top: Expectation values of the different contributions to the total Hamiltonian \eqref{eq:Ht} as a function of the time, with the quench+walk protocol, and with the ergotropy extraction protocol. The values of the control function parameters are taken to be $\gamma_1=4$ and $\gamma_2=1$. For reference, we also plot the ground state energy $E_p^{(0)}$ of $H_p$, Eq.~\eqref{Hp:def} and that, $E_{H(\gamma_2)}^{(0)}$, of the total Hamiltonian $H(\gamma_2)$,  Eq.~\eqref{eq:Ht}. Bottom: fidelity \eqref{fid:def} as a function of time. We have used $N=9$ spins. }
\label{fig1}
\end{figure}

The results for one specific realization of the disorder in \eqref{Hp:def} are shown in Fig.~\ref{fig1} where we plot the energy terms and the fidelity~\eqref{fid:def} as functions of the time.
As a baseline comparison for the efficiency of the quench+walk process, we compare the energy and the fidelity to the corresponding  values obtained for  the ground state of $H_p$ and of the total Hamiltonian $H(\gamma_2)$.

\begin{figure}[h]
\center
\psfrag{ }[ct][ct][1.]{ }
\includegraphics[width=8cm]{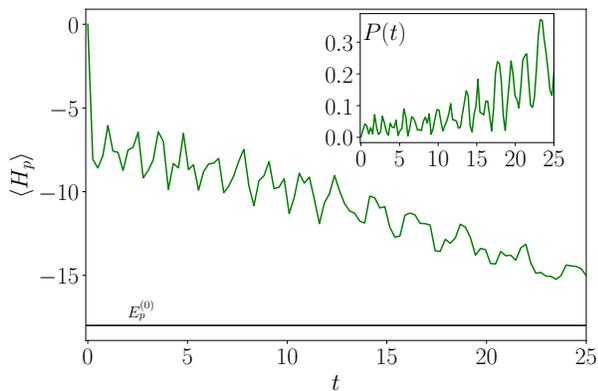}
\caption{Main panel: Expectation value of the problem's Hamiltonian \eqref{Hp:def} as a function of the time, with the annealing protocol  $\gamma(t)=\gamma_1+ (\gamma_2-\gamma_1) t/t_f$   on a system of $N=9$ spins with $\gamma_1=4$, $\gamma_2=1$, $t_f=25$. Inset: fidelity \eqref{fid:def}  as a function of  time. }
\label{fig:ann}
\end{figure}

In Fig.~\ref{fig:ann} we also show the results of an annealing process with a linear control function $\gamma(t)=\gamma_1+ (\gamma_2-\gamma_1) t/t_f$ for the same values of the control parameters $\gamma_1$ and $\gamma_2$ as in Fig.~\ref{fig:ann}. Both figures \ref{fig1} and \ref{fig:ann} will be used as a basis of comparison for the cooling methods introduced and discussed in the following sections.

\nickAdd{We do not consider many of the factors which realistically exist in a large many-qubit annealing system. The reason we have elected to do this is so that our study can isolate the cooling effects from other competing effects which would greatly complicate our analysis. The largest consideration here is that the system will likely be in contact with an uncontrolled bath. These effects are important in real superconducting flux qubit architectures, the only fully controllable annealing systems built with thousands of qubits to date. These effects are not always detrimental, and benefits from interaction with a thermal bath have been experimentally demonstrated~ \cite{dickson13a}. In fact, reverse annealing as implemented on D-Wave systems would not be able to solve optimisation problems without dissipation~\cite{Chancellor2017a,Callison2022a}. Realistically, bath effects will often still be undesirable; it has been demonstrated that coherent regimes can be reached experimentally by quenching very fast~\cite{king2022coherent}, but such approaches are unlikely to be practical in our setting since a large evolution time may be needed. Alternatively, in the future the protocols described here could be simulated on a fault tolerant universal quantum computer, thus avoiding having bath effects altogether. Simulations of annealing protocols on gate-model machines have recently gained interest through the approximate quantum annealing (AQA) algorithm proposed by Willsch et.~al.~\cite{Willsch2022AQA}.}

\section{Computational Cooling}
\label{sec:computational}
In this section we introduce and discuss a computational cooling protocol to  efficiently approach the ground state of $H_p$. 
To this purpose we let the system represented by the Hamiltonian $H_p$  interact with an  auxiliary system which is initially prepared in a state of low energy, see Fig.~\ref{fig:setup}. Specifically we connect the spins of the SK model to an auxiliary 1D transverse-field spin-1/2 Ising chain, whose eigenvalues and eigenstates are well known \cite{Pfeuty1970}.
The total Hamiltonian thus reads 
\begin{eqnarray}
H_{\rm tot}&=&H_p+ \alpha H_A+H_I, \label{Htot:def}\\
H_A&=&-\sum_{i=1}^{N} (1-f) \Sigma^x_{i}\Sigma^x_{i+1}-f \Sigma^z_{i}, \label{HA:def}\\
H_I&=&-J\sum_{i=1}^{N} \sigma^x_{i}\Sigma^x_{i}, \label{HI:def}
\end{eqnarray} 
where we have introduced the Pauli operators  $\Sigma^w_{i}$, $w=x,y,z$ for the auxiliary system, which we assume to be characterised by open boundary conditions ($N+1\equiv 1$ in Eq.~\eqref{HA:def}). The dimensionless parameter $\alpha$ is introduced to lower the value of the ground state energy of the auxiliary system, and as we will see, to enhance the cooling effect when $\alpha>1$.

It is worth  noting that  in the thermodynamic limit the auxiliary system in Eq.~\eqref{HA:def} exhibits a second order QPT at the critical point $f_c=1/2$ (see Ref.~\cite{Pfeuty1970}), with a doubly degenerate ground state and  non-vanishing magnetization along the $x$-axis  for $f<1/2$.

\begin{figure}[t]
\center
\includegraphics[width=8cm]{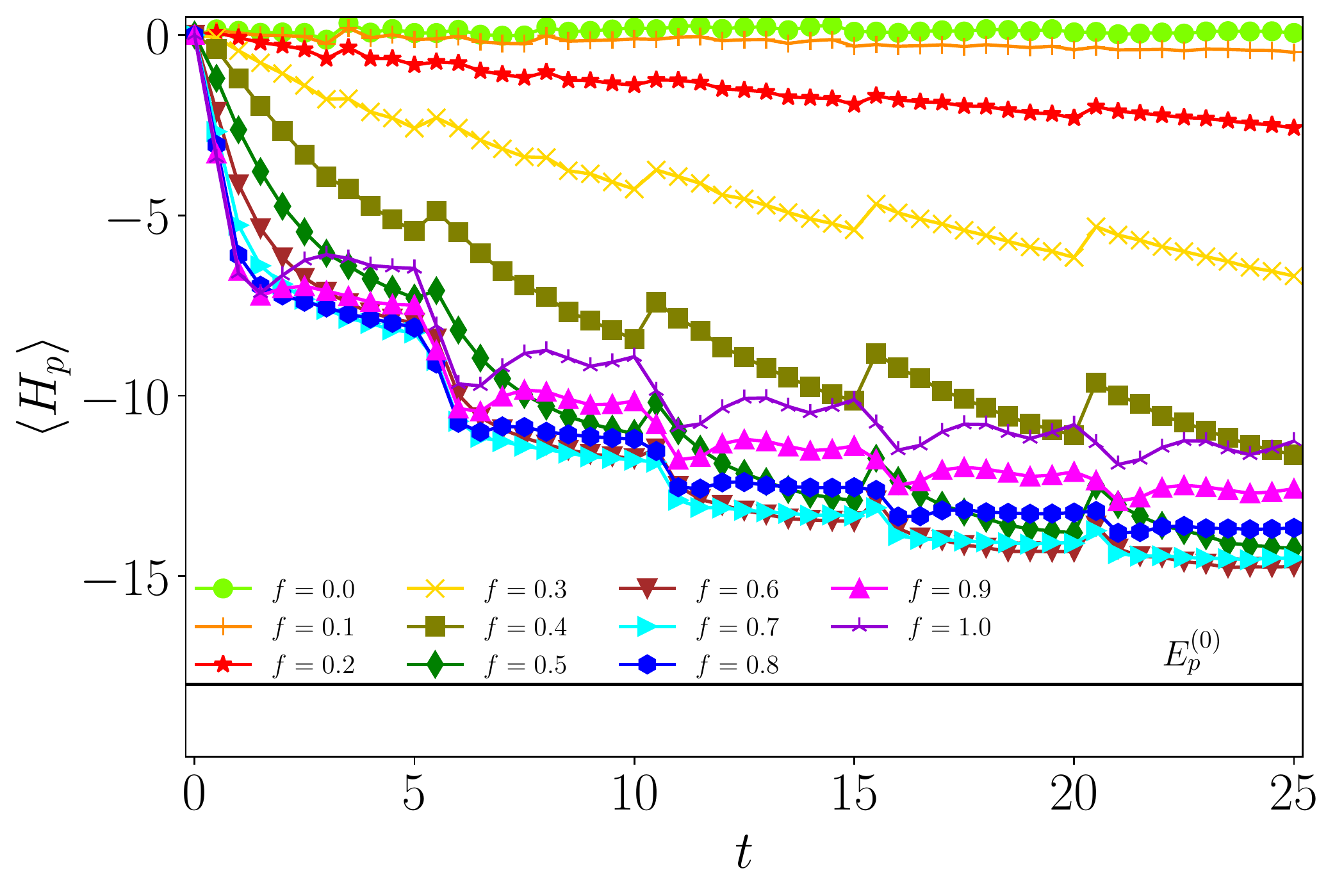}
\includegraphics[width=8cm]{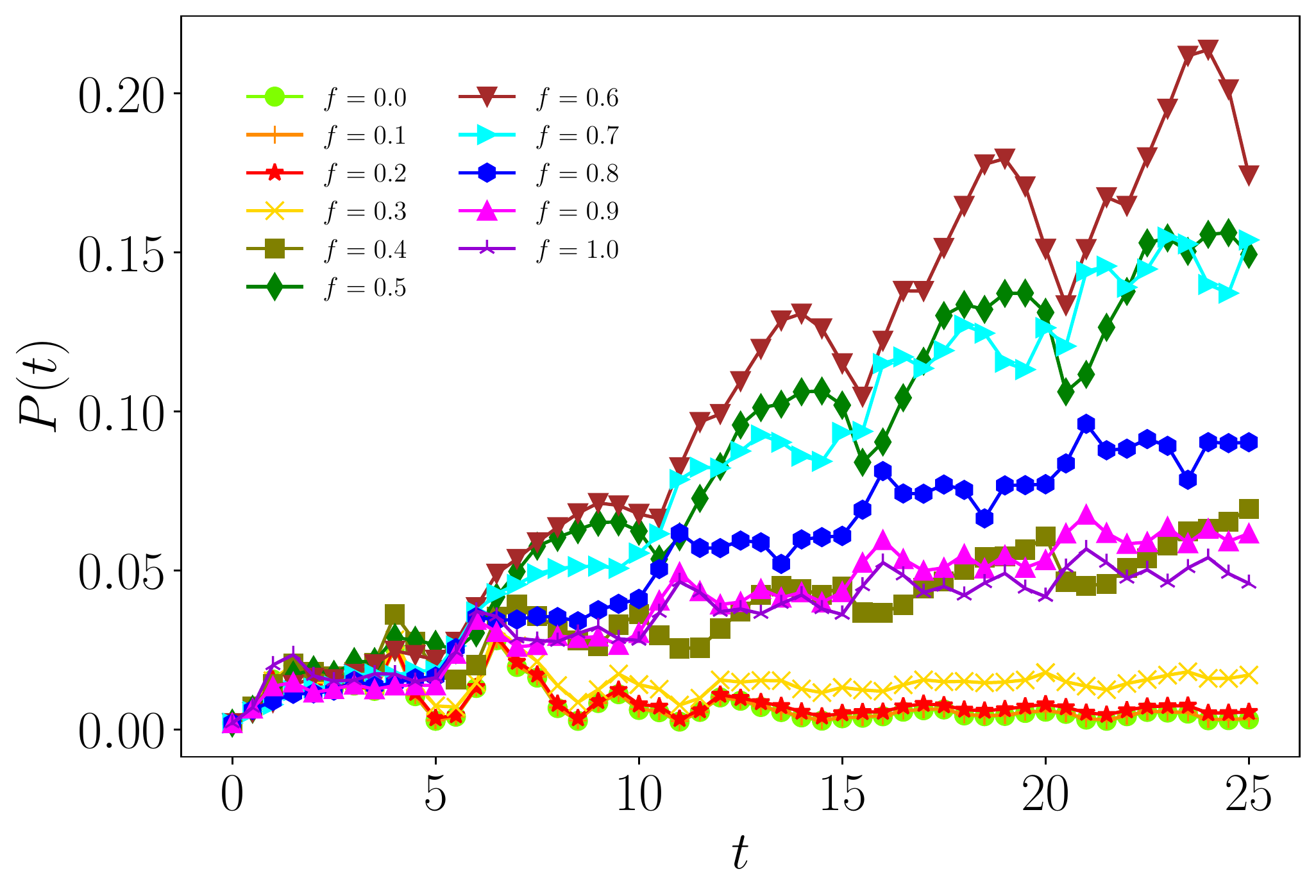}
\caption{ Expectation value of the problem Hamiltonian (top) and fidelity (bottom) as a function of $t$ for the cooling protocol consisting of repeated collisions with the auxiliary system \eqref{HA:def} for different values of $f$,  with $\alpha=3$,  $n_c=5$ cooling strokes of duration $\Delta t=5$ in dimensionless time units. The system is composed of $N=9$ spins, and the parameters of $H_p$ are the same as in Fig.~\ref{fig1}.}
\label{fig:rep:coll}
\end{figure}

We initially prepare the system in the ground state of \eqref{Hd:def}, and the auxiliary system in its own ground state $\ket{E^{(0)}_A}$.
The two systems are then allowed to evolve following the unitary dynamics generated by the total Hamiltonian $H_{\rm tot}$ for a time interval $\Delta t$, after which we disconnect the auxiliary system, and connect the system to a new auxiliary system freshly prepared in its ground state $\ket{E^{(0)}_A}$. We then iterate this procedure for a number $n_c$ of cooling cycles. The resulting problem's energies of this cooling procedure, for different values of the parameter $f$ can be seen in Fig.~\ref{fig:rep:coll}. First of all, we notice that for all $f>0$ we observe some cooling effect in the system. A closer inspection of the results clearly indicates that the cooling procedure is most efficient for $f\simeq 0.6$, both in terms of cooling rate of $\average{H_p}$ and of increasing value of the fidelity.
While the critical value of $f$ for the Hamiltonian $H_A$ is $f_c=1/2$, such a value is renormalized by the interaction with the system through the interaction Hamiltonian, and the value of $f$ required to induce disorder in the auxiliary system becomes larger, $f\simeq0.6$, see the discussion in  Appendix~\ref{app1}. 
We have checked that adding a term of the type $\sigma^y_i\Sigma^y_i$ to the Hamiltonian \eqref{HI:def} does not lead to any substantial change to the results shown in fig.~\ref{fig:rep:coll}.

Comparing the results of Fig.~\ref{fig:rep:coll} with those of Fig.~\ref{fig1}, we see that that connecting the system to the external cooler improves significantly the optimization of the problem's energy. The lowest energy obtained with the repeated collisions are similar to that obtained with the annealing reported in Fig.~\ref{fig:ann} in a similar timescale. 

\begin{figure}[h]
\center
\psfrag{ }[ct][ct][1.]{ }
\includegraphics[width=8cm]{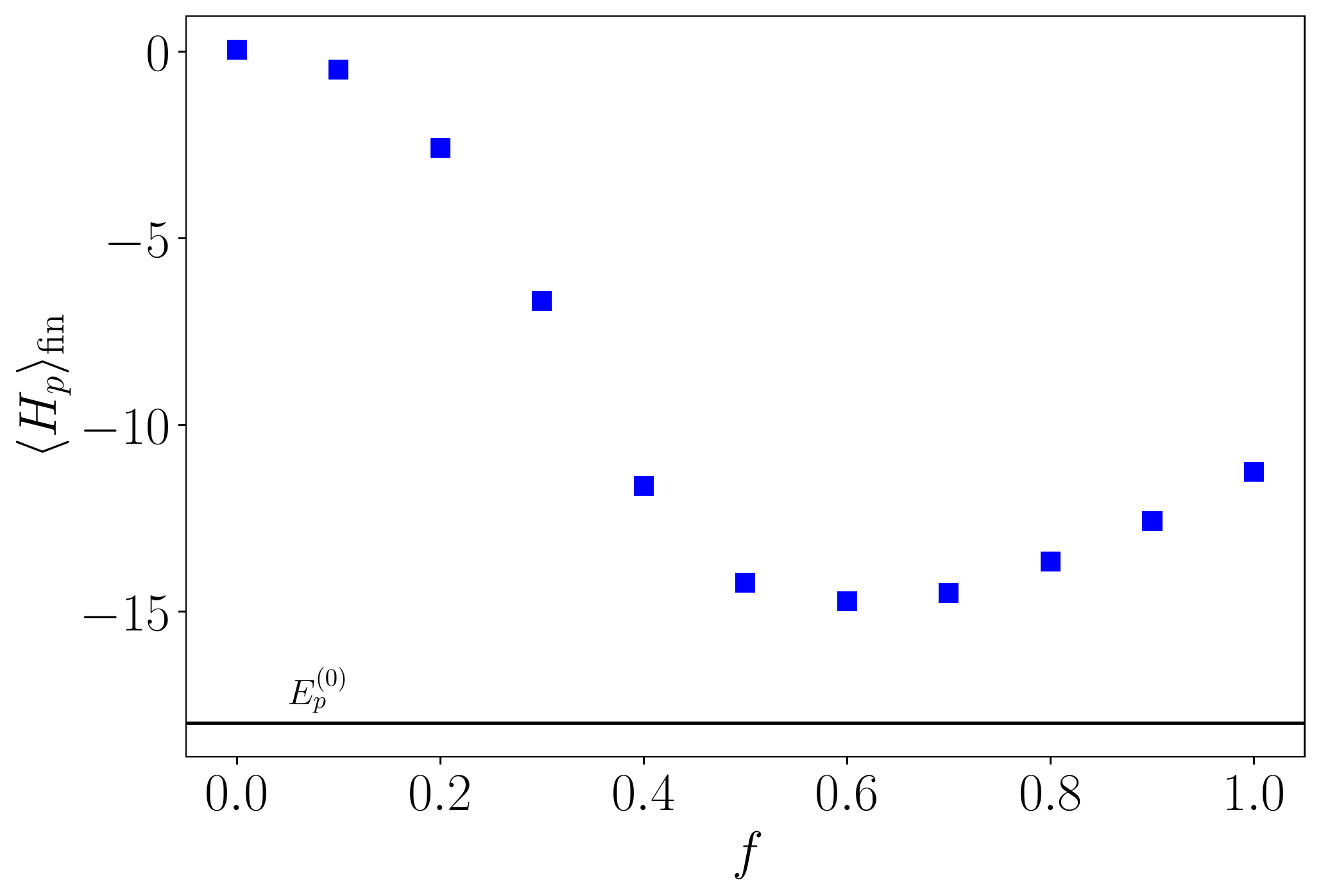}
\caption{Final value of $\average{H_p}$ as a function of $f$.  For each value of $f$,  $\average{H_p}_{\mathrm{fin}}$ is the final value of the system energy at the end of the last cooling stroke of Fig.~\ref{fig:rep:coll}.}
\label{fig:coll:fin}
\end{figure}

We also evaluate the long time behaviour of  $\average{H_p}$ as a function of $f$, as given by the final values of $\average{H_p}$ at the end of the last cooling stroke. The results are shown in Fig.~\ref{fig:coll:fin}: inspection of this figure confirms that the maximal cooling effect is obtained for $f\simeq 0.6$.

In Appendix \ref{app:B}, we complement our study by 
investigating  the effect of the parameter $\alpha$ renormalizing the auxiliary system Hamiltonian (see Eq.~\eqref{Htot:def} on the cooling protocol. Furthermore, we consider the effect of  {\it freezing} the system dynamics by reducing the interaction strength $J$ appearing in Eq.~\eqref{HI:def}. We also show that the long time behaviour of  the system undergoing repeated collision with the auxiliary bath is  practically independent of the chosen initial state.   
\alb{Finally, in the same Appendix, we consider different realizations of the quenched disorder in the SK chain (different sets of values of $J_{ij}$ and $h_i$): the cooling performance of our protocol does not show any substantial change, see Fig.~\ref{fig:seed}.}

\begin{figure}[h]
\center
\includegraphics[width=8cm]{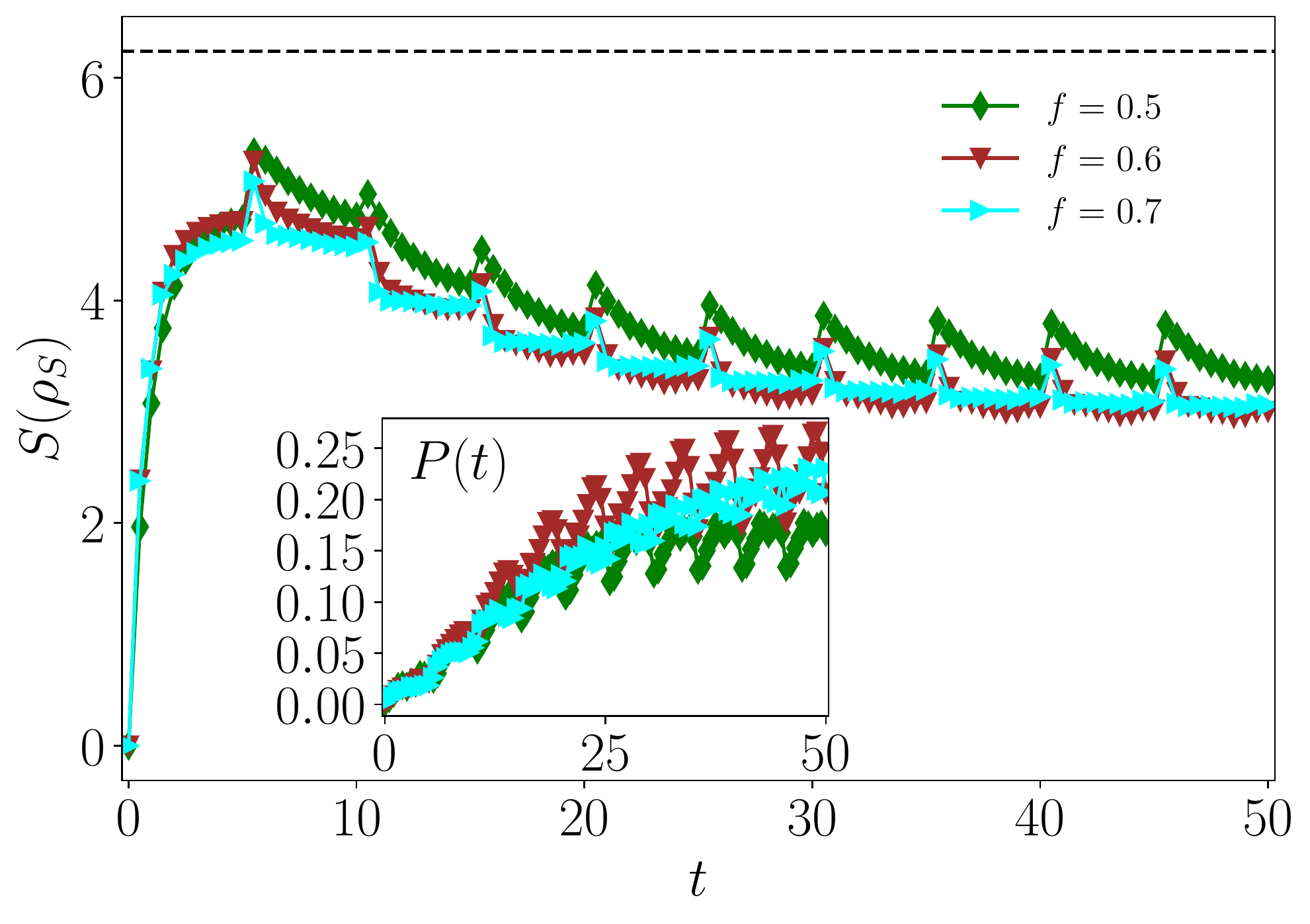}
\caption{ Main panel:  von Neunmann entropy of the reduced density matrix of the system $S(\rho_S)$ as a function of the time for three values of $f$. The dashed line corresponds to the entropy of the completely random state $N\ln 2$. Inset: fidelity as a function of the time for the same values of $f$.  $\alpha=3$,  $n_c=10$ cooling strokes of duration $\Delta t=5$ in dimensionless time units. The system is composed of $N=9$ spins, and the parameters of $H_p$ are the same as in Fig.~\ref{fig1}.}
\label{fig:entro}
\end{figure}

In order to evaluate the purity of the system across different strokes, we compute the von Neumann entropy of the reduced density matrix of the system:
\begin{equation}
    S(\rho_S) = -{\rm tr} \rho_S \ln \rho_S. 
\end{equation}
In Fig.~\ref{fig:entro}, we report $S(\rho_S)$
as a function of the time. After the first cycle the system gets close to the maximally mixed state, corresponding to a large system-environment entanglement. 
Subsequently, $S(\rho_S)$ decreases as the number of cooling cycles increases.
Thus, while the repeated collision protocol reduces the energy of the system, such a cooling is achieved thanks to the establishment of quantum correlations with the auxiliary system. This was also noted in smaller collision models \cite{DeChiaraPRR2020}.
In the inset of the same figure we plot the fidelity as a function of the time for ten cooling cycles.

\alb{It is worth noticing that the process describing the evolution of the system state is non--Markovian. Indeed we use a finite interaction time $\Delta t=5$ in dimensionless units which does not lead to a Gorini-Kossakowski-Susarshan-Lindblad (GKSL)  master equation for the system of interest. The evolution of an open quantum system is Markovian if and only if the process is described by a GKSL  master equation, see, e.g. \cite{Rivas_2014} and references therein. The GKSL master equation is retrieved if one considers vanishing  interaction time $\Delta t$ between the bath and the system, and interaction strength that scales as $H_I\sim1/\sqrt{\Delta t}$ \cite{De_Chiara_2018}. In Appendix \ref{app:B} we investigate the performance of the cooling protocol in that regime and find that the finite time interaction protocol performs considerably better in a single interaction stroke than the Markovian one with several short collisions. Thus we conclude that the bath and the system need to interact for a finite time for the bath to extract an appreciable amount of heat from the system.
}

\albb{Quantum thermalization machines made of ancillary spins were also proposed in \cite{Arisoy2019,Arisoy2021} to realise the thermal states of single spins or finite many-body systems. In those references, the authors also used repeated interactions with engineered environments, however a fine tuning of the baths energy level was required to achieve the target thermal states.}

 \nickAdd{It is worth briefly discussing the practical implementation of the protocol depicted in Fig.~\ref{fig:setup}. The most naive approach is to consider a new spin chain being coupled for each cooling cycle. Such an implementation would be impractical, since the number of qubits would grow linearly with the number of cooling cycles. Alternatively, if we had a separate procedure to ``reset'' the same spin chain to an approximate ground state, we could reuse the same chain, see for instance Ref.~\cite{MeloPRA2022}. This would introduce a period of time during which the chain was reset and the spin glass was allowed to undergo dynamics. Based on numerical observations in \cite{Callison2019}, it is likely that during this time period the system would rapidly equilibrate toward a long-time average, so adding such a time period would be unlikely to be detrimental. Adding such a time period would however complicate our model and give us another parameter to consider so it is therefore undesirable at the level of detail which we are working. We argue that the cooling time could be avoided using a constant number of spin chains using the following procedure: after each spin chain has interacted with the system it immediately begins the resetting procedure, meanwhile another already reset spin chain interacts with the system. As long as the number of spin chains is greater than the ratio of the resetting time to the time of each cycle, a reset spin chain will always be available, see also discussion in Ref.~\cite{PiccionePRA2021}.}

\nickAdd{One question is how the coupling could be turned on or off in practice, but this will depend on the underlying system. For example, in circuit based flux qubit architectures, the coupling elements are the same as the qubits, but biased into a monostable rather than bistable regime \cite{den_Brink2005coupler}. In such an architecture it would be relatively straightforward to tune coupling by independently adjusting the biases on the different qubits (and single-body correction factors on the qubits they couple). In this setting a model of individually tuning couplers is realistic, although instantaneously turning off coupling may be difficult, making a model where the coupling is reduced non-instantaneously over time slightly more realistic. Simulating such systems would however be much more numerically demanding. Since we are interested in qualitative behaviour of realistic systems rather than simulating real devices, instantaneous switching is sufficient for these purposes. Likewise, in atomic systems, coupling could be controlled by manipulating the physical separation between the system and bath qubits could be an effective method of turning coupling off, especially since all couplings are turned on and off at the same time. Shuttling of ions is considered to be an important component to trapped ion quantum computing and protocols to effectively and quickly move ions are being developed \cite{Kaushal2020shuttling}, so this could be a realistically achievable way of controlling the coupling.}

\nickAdd{As for the resetting procedure itself, we first note that, strictly speaking, the third law of thermodynamics forbids a system from being prepared exactly in its ground state in a finite time. For our purposes however, we assume that the spin chain is prepared in an approximate ground state with a finite energy gap, where, at a small but finite temperature, probabilities to be excited can be neglected. In fact a common approximation in quantum annealing is to assume that the system originates in the ground state of the driver Hamiltonian~\cite{Albash2018Adiabatic}, in many ways our approximation is no different. As this state is an approximate thermal state, it is a free thermodynamic resource, see for instance Ref.~\cite{Lostaglio_2019}.}
\gab{From such a simple product state, it is possible to prepare strongly correlated systems that are ground states of quantum spin models, Ref.~\cite{VerstraetePRA2009,Cervera2018}. This would require a number of gates that scales polynomially with the number of qubits in the environmental chain. This gates may require external thermodynamic work to be realised which therefore scales polynomially with the number of qubits.}

\gab{We emphasise that our proposal can also be implemented in quantum circuit models as a quantum collision model, also known as a repeated interaction model. In fact, small quantum collision models have already been demonstrated in superconducting quantum circuits~\cite{MeloPRA2022, CattaneoPRXQ2023}}
 
\section{Cooling by measuring the auxiliary system}
\label{sec:measuring}
In this section we consider again the system described in Eqs.~\eqref{Htot:def}--\eqref{HI:def} and combine the collisional cooling method discussed in the previous section, with projective measurements on the auxiliary system.
Let $\ket{\EAj}$ be the eigenstates of $H_A$ with projector $\Pi^{(j)}_{A}=\ket{\EAj}\bra{\EAj}$, and  $\rho_{\rm tot}$ the state of the combined  system plus environment at a given time. The probability  of  measuring the energy $\EAj$ is then given by
\begin{equation}
p(\EAj)=\tr[ \Pi^{(j)}_{A} \rho_{\rm tot}], 
\label{eq:pAj}
\end{equation} 
and  the post-measurement reduced state of the system is 
\begin{equation}
\rho_S(t|\EAj)= \tr_A [ \Pi^{(j)}_{A}\rho_{\rm tot}]/p(\EAj). \label{eq:rhosAj}
\end{equation} 
One can then evaluate the expectation value of $\average{H_p}$
  conditioned by the outcome of  measurement $\EAj$ on the auxiliary system
\begin{equation}
\average{H_p}_{\EAj}= \tr_S [\rho_S(t|\EAj) H_p].
\label{Hp:cond}
\end{equation}


\begin{figure}[t]
\center
\psfrag{ }[ct][ct][1.]{ }
\includegraphics[width=8cm]{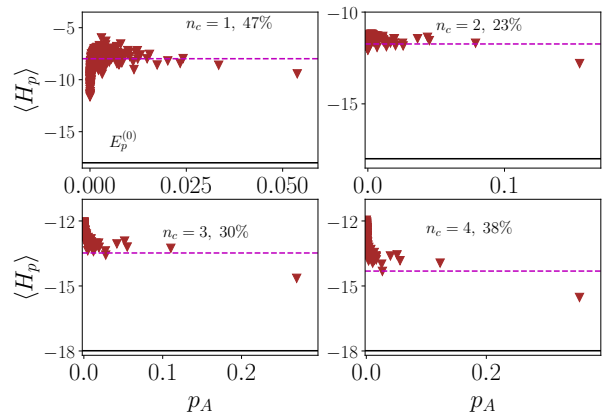}
\caption{Post-measurement system energy \eqref{Hp:cond} as a function of $p_A$ as obtained by measuring $H_A$ at the end of the first four cooling strokes, for different values of $f$, 
 with $\alpha=3$ and $N=9$. The rest of parameters are as in Fig.~\ref{fig:rep:coll}. Dashed lines: $\average{H_p}$ at the end of the cycle, i.e. immediately before the measurement, see also Fig.~\ref{fig:rep:coll} for the cycles with $\alpha=3$. The percentage gives the fraction of measurements that return a value of $\average{H_p}$ below the value at the end of the cycle. }
\label{fig:meas:4cycles}
\end{figure}


To evaluate the effect of a measurement after the repeated cooling strokes, we  first apply the measure protocol to the cooling dynamics as depicted in Fig.~\ref{fig:rep:coll}: at the end of a given number of cooling cycles we evaluate the probability $p(\EAj)$ of measuring a given eigenstate of $H_A$, and the corresponding value of the post-measurement system energy  $\average{H_p}_{\EAj}$. More precisely, we set $n_c=1,\,2, \, 3$ or 4, cool the system as described in the previous section, and then measure the energy of $A$.  The results of such a measurement protocol are shown in Fig.~\ref{fig:meas:4cycles}, for different  values of the parameter $f$.
We see that,  in general,  measuring $H_A$  has  no practical beneficial effect in the cooling of the system: the total probability that a measurement returns a value of   $\average{H_p}$ lower that the value at the end of the cooling stroke is smaller than 50\%. However, at the end of the 4th stroke and for $\alpha=2,\, 3$ there is a significant probability ($\gtrsim 40$\%) that the measurement 
returns a value of  $\average{H_p}$ lower than the one at the end of the cooling stroke, resulting in additional net cooling.

Next, we  investigate whether the measurement of the auxiliary system can be used to speed-up the cooling of the system of interest and enhance the fidelity with respect to the target state Eq.~\eqref{fid:def} when compared to the simple collisional cooling shown in Fig.~\ref{fig:rep:coll}.

To this end, we implement the following cooling algorithm: at $t=0$ we prepare both the system of interest and the auxiliary system as described above, i.e., the system starts in the ground state of \eqref{Hd:def} while the auxiliary cooler $A$ starts in its own ground state. We then let the composite system evolve with the dynamics generated by the total Hamiltonian \eqref{Htot:def} for a time $\Delta t$, at the end of which we perform a measure of $H_A$, calculate the corresponding system post-measurement reduced state \eqref{eq:rhosAj} and start a new cycle by connecting the system with  a  new auxiliary system freshly prepared in its ground state $\ket{E^{(0)}_A}$. We then iterate the above procedure for a given number of strokes.
We consider two different protocols where we combine collisional cooling and measurements.

\begin{figure}[t]
\center
\includegraphics[width=8cm]{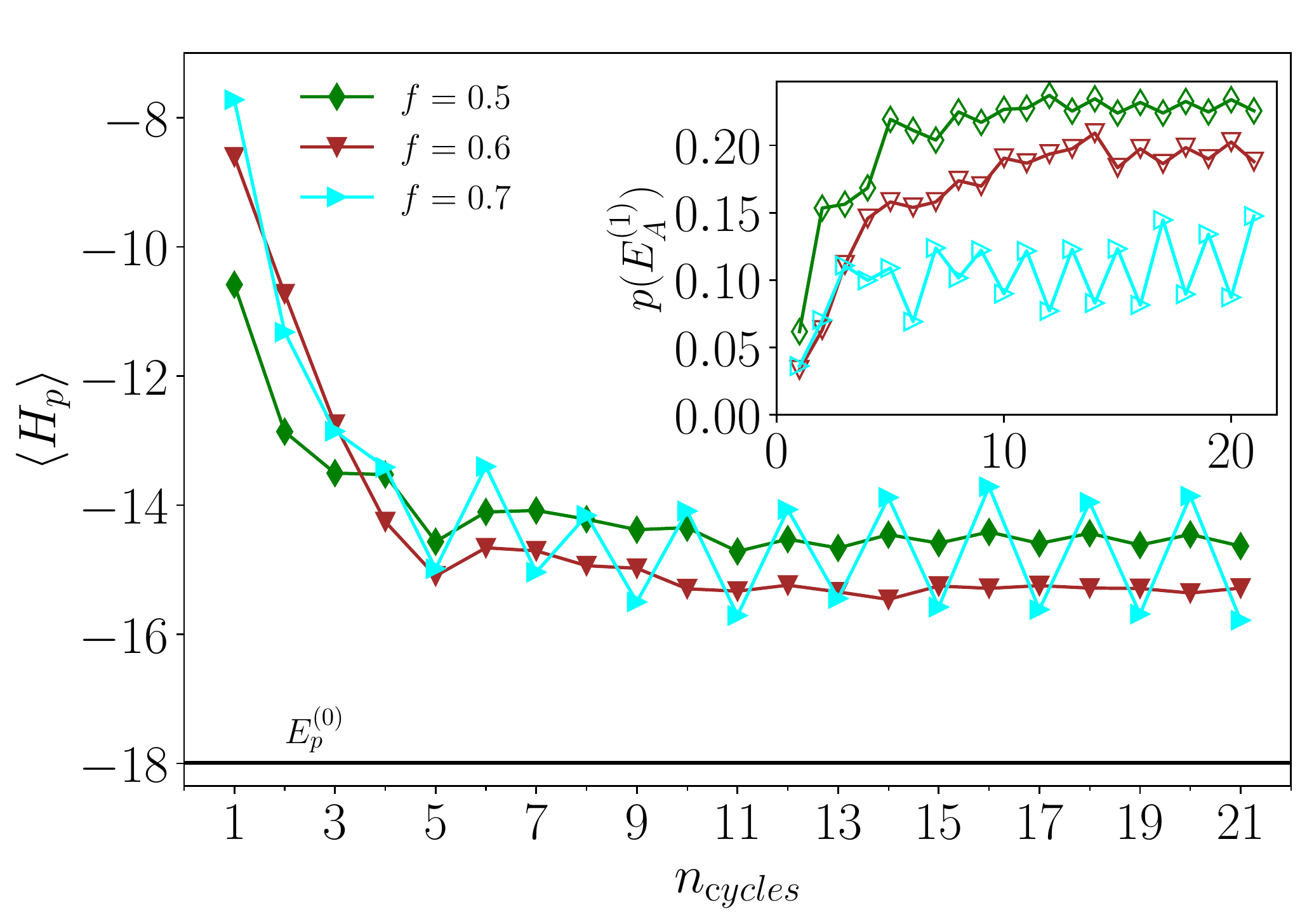}
\includegraphics[width=8cm]{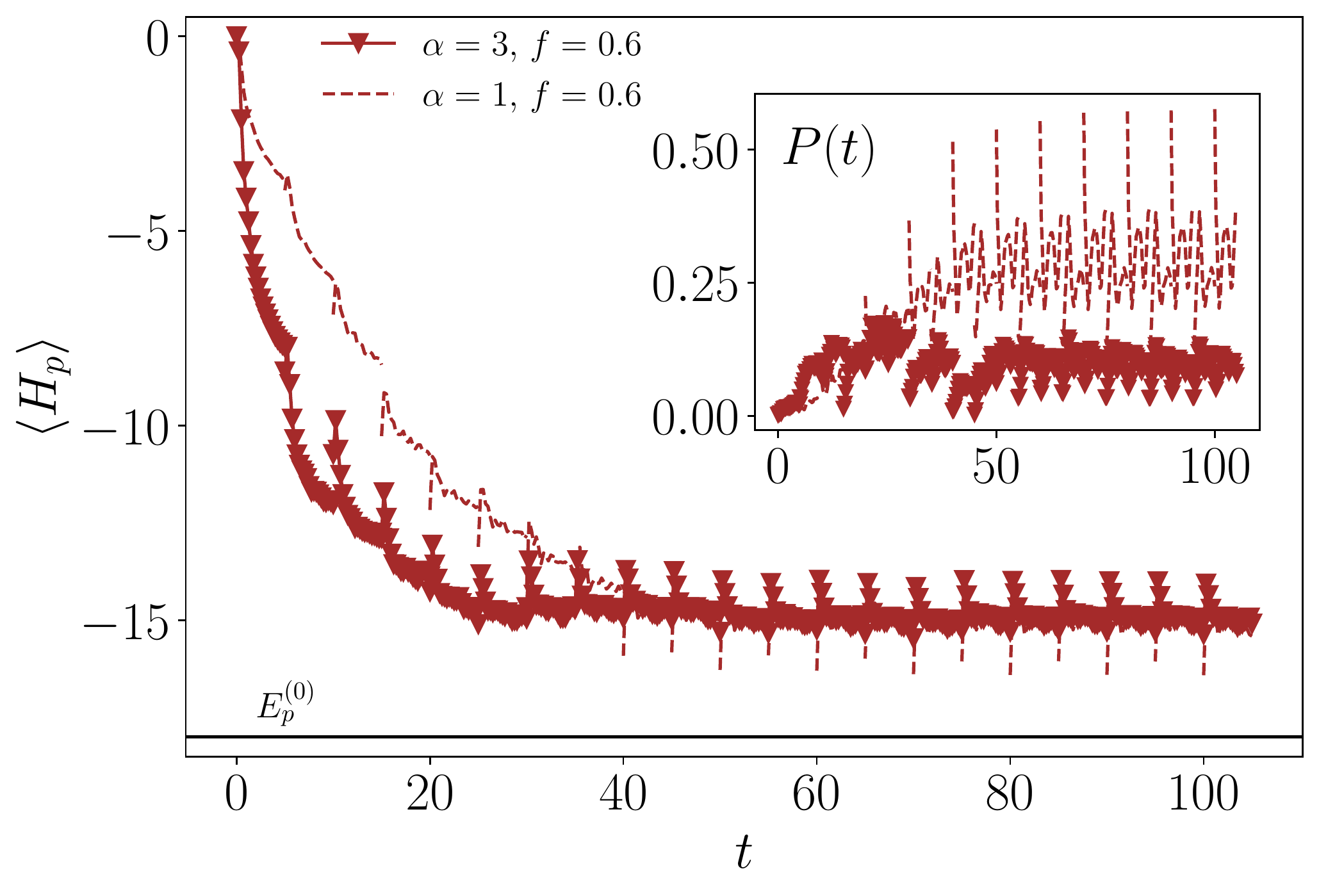}
\caption{Collisional cooling strokes followed by a projective measurement of the auxiliary system Hamiltonian $H_A$ with $\alpha=1$, $\Delta t=5$  and $J= 1$ (see Eqs.~\eqref{Htot:def}-\eqref{HI:def}).
In this protocol we select the post-measurement state corresponding to the second smallest eigenvalue of $H_A$: $E^{(1)}_A$. Top panel:  expectation value of the system Hamiltonian $H_p$ after the measurement as a function of the number of cooling cycles.  Inset: probability of the selected state of the auxiliary system $p_A(E_A^1)$. Bottom panel: expectation value $\average{H_p}$ as a function of $t$. Inset: fidelity as a function of $t$. }
\label{fig:rep_meas}
\end{figure}

First, we only select post-measurement states corresponding to the second smallest eigenvalue of $H_A$, $E_A^{(1)}$,
 see Fig.~\ref{fig:rep_meas}. The rationale behind  this choice is that, being the $A$ system prepared in its ground state, finding it in its first excited state is the minimal requirement for the system of interest to be cooled. While this specific protocol might be of limited practical use, as the probability of achieving a sequence of exactly $n_c$   measurements with $E_A=E_A^{(1)}$ can be pretty small, it is however a useful exemplification of the cooling mechanism obtained by energy transfer to the $A$ system prepared in a {\it cold} state.
This scenario might however be interesting in presence of a Maxwell daemon,  selecting only the post-measurement state $\ket{ E_A^{(1)}}$. The feasibility of this setup and its thermodynamic cost is however beyond the scope of the present paper.

\begin{figure}[t]
\center
\psfrag{ }[ct][ct][1.]{ }
\includegraphics[width=8cm]{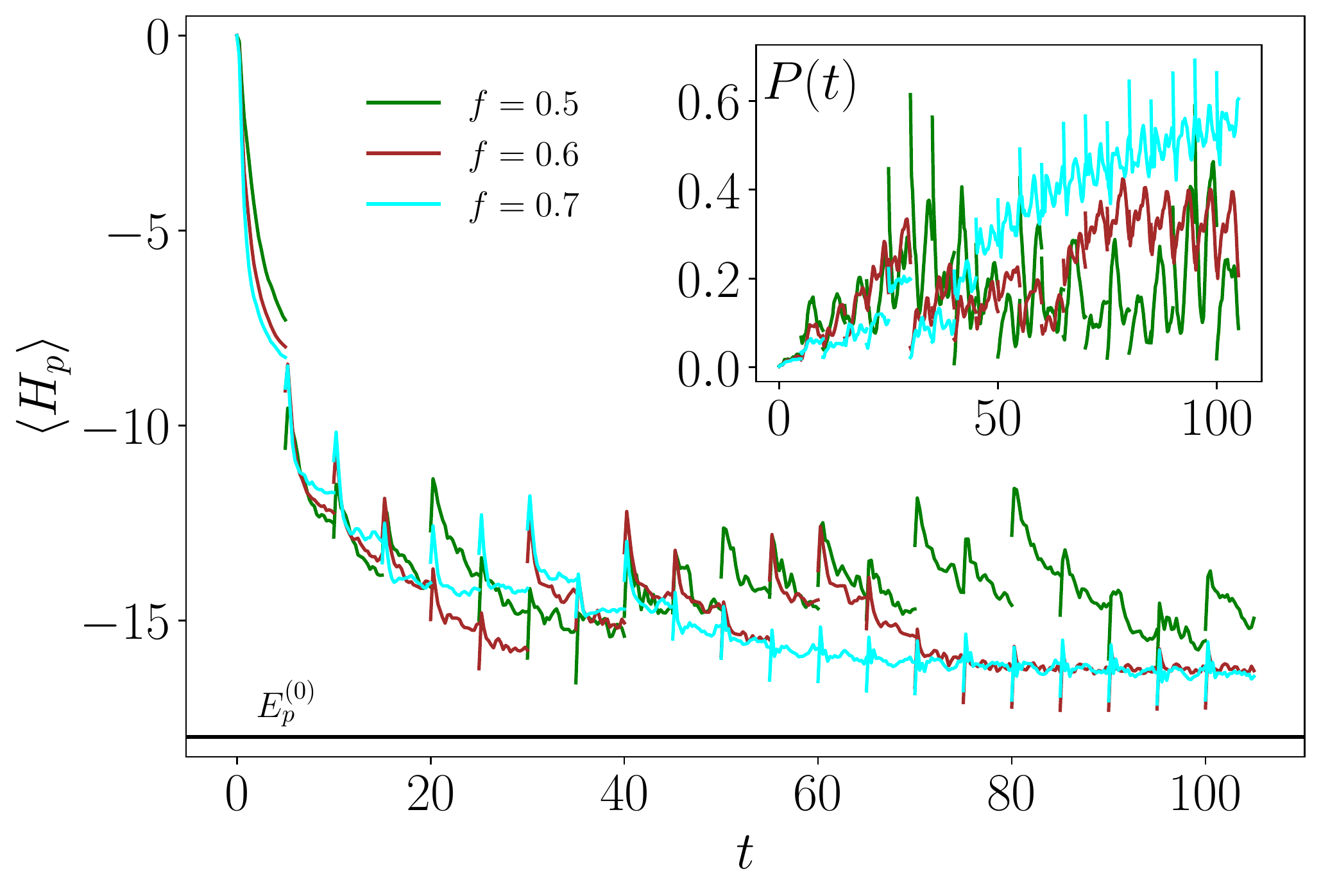}
\includegraphics[width=8cm]{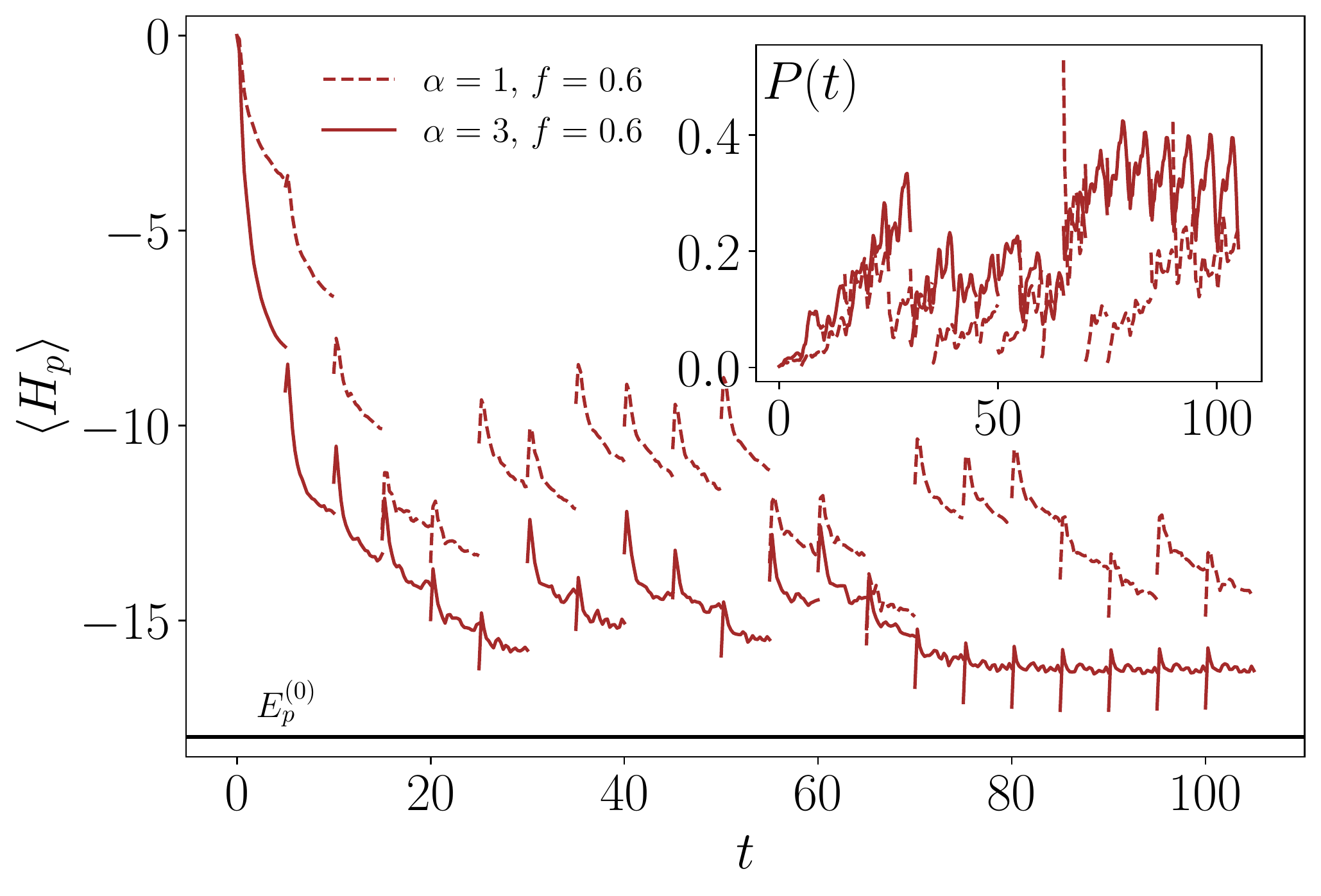}
\caption{Collisional cooling strokes followed by a projective
measurement of the auxiliary system Hamiltonian $H_A$  (see eqs.~\eqref{Htot:def}--\eqref{HI:def}).
In this protocol we randomly select the post-measurement state with probability $p(E^{(j)}_A)$, Eq.~\eqref{eq:pAj}. Top figure, main panel: expectation value $\average{H_p}$ as a function of $t$ for a single realization of the repeated measurement process. Inset: fidelity as a function of $t$. Bottom figure: $\average{H_p}$ as a function of $t$  for 2 different values of $\alpha$ and $f=0.6$. }
\label{fig:rep_meas_MC}
\end{figure}

Second, we consider a more viable protocol consisting of a stochastic post-measurement selection protocol, more similar to a realistic experimental setup. At the end of each  cooling stroke, we perform a measurement of $H_A$ and as a consequence the system collapses in a post-measurement state $\rho_s(t|\EAj)$ with probability $p(\EAj)$, see Eqs.~\eqref{eq:pAj}--\eqref{eq:rhosAj}.
After such a projective measurement,  we disconnect the system from the $A$ system, and connect it to a new $A$ system prepared  in its ground state and iterate the procedure.
The corresponding results are shown in Fig.~\ref{fig:rep_meas_MC}. The curves in such a figure should be compared with those in Fig.~\ref{fig:rep:coll} and ~\ref{fig:rep:comp} obtained with repeated collisions alone for the same value of the parameter $f$. Combining collisions and repeated measurements gives a similar decreasing rate for  $\average{H_p}$. Averaging over many stochastic ``trajectories'' as the one depicted in Fig.~\ref{fig:rep_meas_MC}, one would obtain the curves for  $\average{H_p}$ and $P(t)$ shown in Fig.~\ref{fig:rep:coll}, which thus represent the expectation values of the quantities under scrutiny when repeating the stochastic post-measurement selection protocol several times.

\section{Conclusions}
\label{sec:conclusions}
In this paper we have presented a thermodynamics-inspired protocol to find the minimum energy  of quantum systems characterized by complex Hamiltonians.
Specifically we have considered a classical optimization problem, that is turned into a quantum cooling process by using a non-Markovian bath that induces coherent cooling dynamics in the system of interest.
The bath is initially prepared in its ground state and lowers the system energy through repeated collisions.
The optimal working regime for our machinery is found in the range of parameters where the bath exhibits a QPT in the thermodynamic limit, thus emphasizing the importance of collective effects in thermal devices.

We  conjecture that this result is not restricted to the specific bath considered in the present paper: any bath operating in its critical region should be extremely ``susceptible" to external disturbances and thus more capable of exchanging energy from the systems they are put in contact with.

By combining the cooling protocol with measurements on the bath alone, we are able to improve further the protocol, and in particular to keep the post-measurement state of the system pure.
The possibility of using feedback control, mimicking a Maxwell daemon, with post-measurement selection rules affecting the bath alone seems to be an interesting direction for future explorations.

\nickAdd{It would be interesting to compare to other algorithms, for example gate model variational algorithms like the quantum approximate optimization algorithm \cite{Callison2022a, Farhi2022}. Due to the similar continuous-time setting we have limited the present study to comparison with annealing and multi-stage quantum walks, but a broader comparison would be enlightening. A systematic analysis of the thermodynamic and computational resources required for our protocol and the comparison with alternative schemes remains an open problem.}

While we studied the performance of the protocol in terms of cooling and fidelity rate for a wide range of parameters, it can be further improved by systematic optimization of the relevant parameters. In particular it might be interesting to consider baths with a number of spins different from the system of interest. 

\acknowledgements
We acknowledge the support by the UK EPSRC EP/S02994X/1 and EP/T026715/1, and the Royal Society IEC\textbackslash R2\textbackslash 222003.

A.I. gratefully acknowledges the financial support of
The Faculty of Science and Technology at Aarhus University through a Sabbatical scholarship and the hospitality of the Quantum Technology group, at the School of
Mathematics and Physics, Queen's University Belfast, during the initial stages of this project.

\appendix
\section{Effect of $f$ on the static properties of the $A$ system}
\label{app1}
\begin{figure}[h]
\center
\psfrag{ }[ct][ct][1.]{ }
\includegraphics[width=8cm]{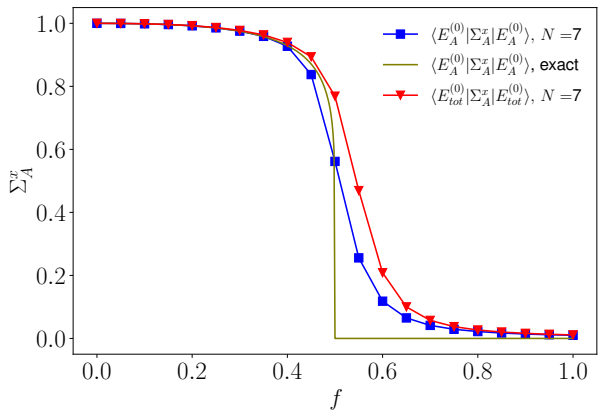}
\caption{$x$--magnetization of the $A$ system in presence or absence of the interaction with the problem Hamiltonian. The full line is the exact solution for $\average {\Sigma^x}$ in the thermodynamic limit \cite{Pfeuty1970}.}
\label{fig:sig}
\end{figure}

In the main text we have seen that the optimal cooling effect obtained with repeated collisions with the   $A$ system is obtained for $f\simeq0.6$, see Eqs.~\eqref{Htot:def}-\eqref{HI:def}. In this Appendix we compare the behaviour of the auxiliary system with  ($J\neq0$ in Eq.~\eqref{HI:def}) or without interaction ($J=0$) with the problem Hamiltonian.
We remind the reader that the spectrum and the eigenstates of Hamiltonian \eqref{HA:def} have been fully characterized in \cite{Pfeuty1970}.
In particular, the ground state of $H_A$ becomes doubly degenerate for $f<0$ with non--vanishing magnetization along the $x$-axis for $f<f_c=1/2$.
In Fig.~\ref{fig:sig} we compare the expectation value of $\Sigma^x$ in presence or absence of interaction with the problem Hamiltonian for a system with $N$ spins, together with the exact solution of the magnetization in the thermodynamic limit. We see clearly that the interaction with the problem Hamiltonian shifts the value of $f$ for which the auxiliary system exhibits disorder toward a larger value than $f_c$. This must be compared again with the dynamical results of the collisional cooling, as exemplified in, e.g., Fig.~\ref{fig:rep:coll} or \ref{fig:coll:fin}, where we show very clearly that the optimal cooling effect is obtained for values of $f$ larger than $f_c=1/2$.
Thus a comparison of  Fig.~\ref{fig:sig} with Fig.~\ref{fig:rep:coll} and  \ref{fig:coll:fin} suggests that the optimal cooling regime occurs for values of $f$ for which an order-disorder transition occurs in the auxiliary system coupled to the system of interest.

\section{Additional results}
\label{app:B}

\begin{figure}
\center
\includegraphics[width=8cm]{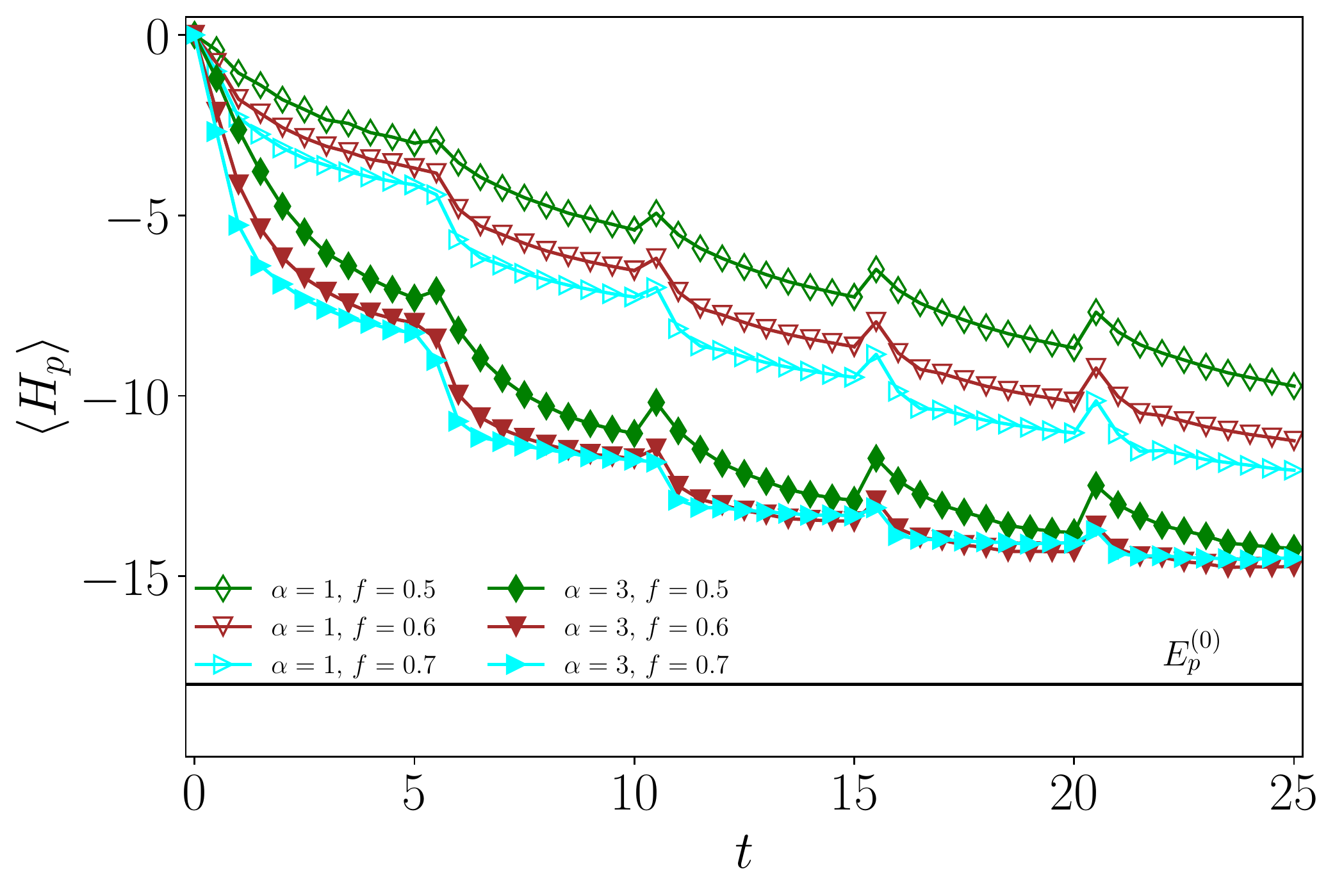}
\includegraphics[width=8cm]{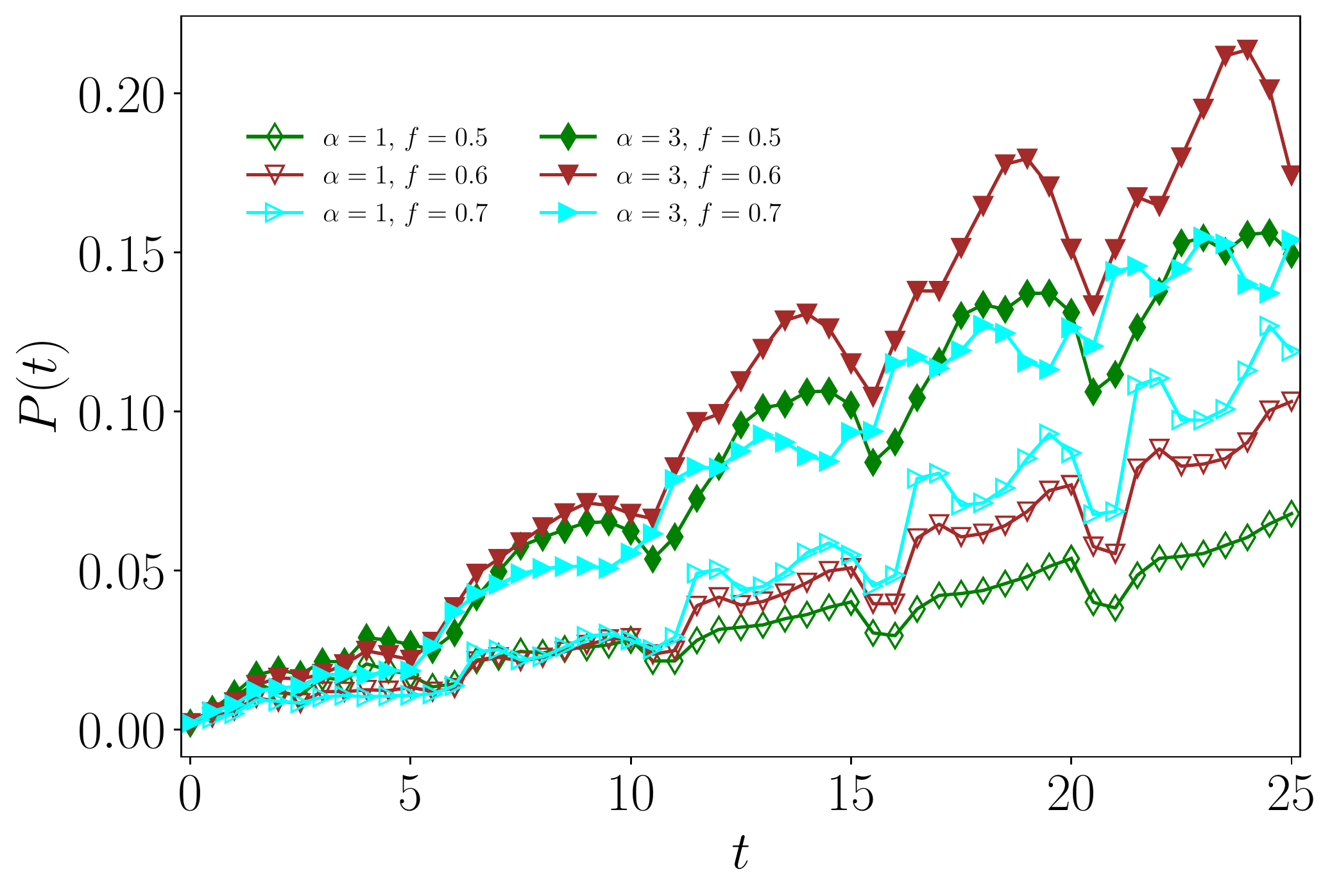}
\caption{Expectation value of the problem Hamiltonian (top) and fidelity (bottom) as a function of $t$ for the cooling protocol consisting of repeated collisions with the auxiliary system \eqref{HA:def} for different values of $f$ and $\alpha$, with  $n_c=5$ cooling strokes of duration $\Delta t=5$ in dimensionless time units. The system is composed of $N=9$ spins, and the parameters of $H_p$ are the same as in Fig.~\ref{fig1}. Larger values of $\alpha$, e.g. $\alpha=4$, results in a deterioration of the cooling performance (data not shown).}
\label{fig:rep:comp}
\end{figure}

In Fig.~\ref{fig:rep:comp} we study the effect of the parameter $\alpha$ renormalizing the auxiliary system Hamiltonian (see Eq.~\eqref{Htot:def}) on the cooling protocol. We see that increasing $\alpha$ improves the performance of the protocol introduced in this section, both in terms of the lowest problem's energy reached and the time to reach it. This is due to the fact that, for larger $\alpha$, the system sees the auxiliary bath as  {\it colder}, given that we initially prepare it in its ground state $\ket{E^{(0)}_A}$ as discussed above.

\begin{figure}
\center
\includegraphics[width=8cm]{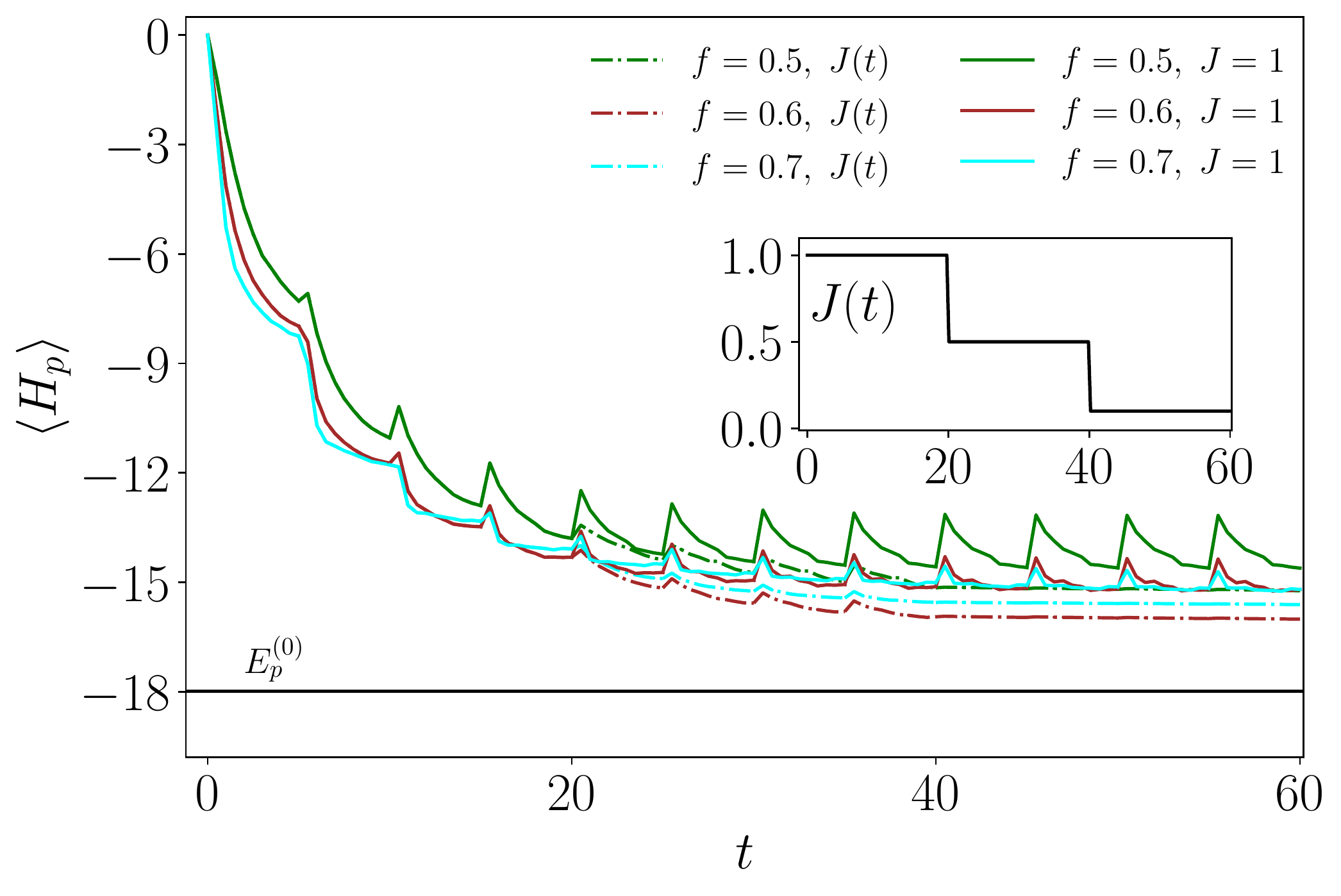}
\includegraphics[width=8cm]{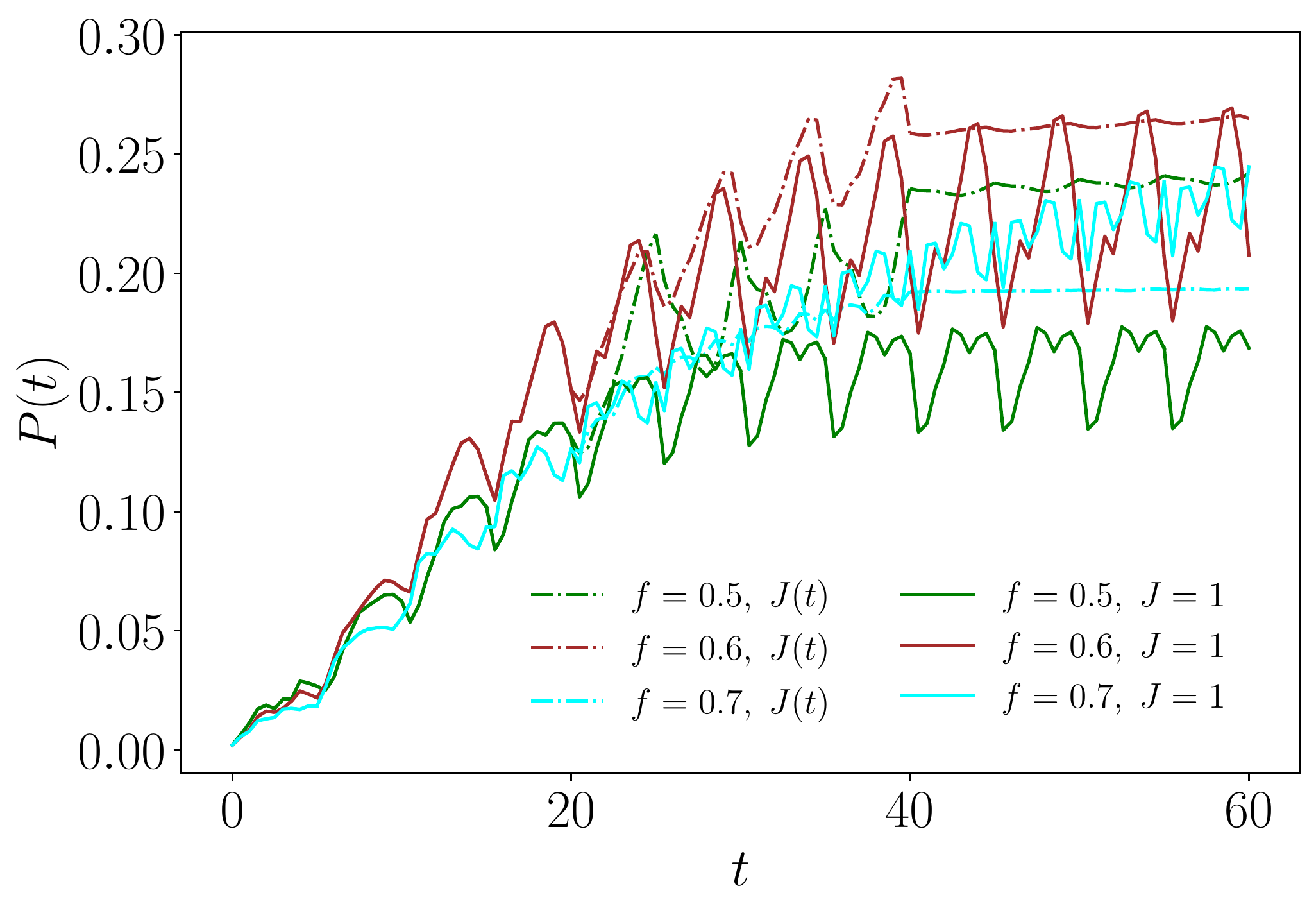}
\caption{ Expectation value of the problem Hamiltonian (top) and fidelity (bottom) as a function of $t$ for the cooling protocol consisting of repeated collisions with the auxiliary system \eqref{HA:def} for different values of $f$,  with $\alpha=3$,  $n_c=12$ cooling strokes of duration $\Delta t=5$ in dimensionless time units. The system is composed of $N=9$ spins, and the parameters of $H_p$ are the same as in Fig.~\ref{fig1}. The dashed lines correspond to a time protocol where we perform a quench on $J$ every four strokes, the continuous line correspond to constant interaction strength  $J=1$. The time protocol $J(t)$ is plotted in the inset of the first figure.}
\label{fig:reduceJ}
\end{figure}

In Fig.~\ref{fig:reduceJ} we study the effect of {\it freezing} the system dynamics by reducing the interaction with the auxiliary system: specifically we implement two subsequent quenches on $J$, i.e., the interaction strength appearing in Eq.~\eqref{HI:def}. Inspection of Fig.~\ref{fig:reduceJ} confirms that decreasing the interaction strength between the system and the auxiliary cooler simply freezes the system in its state.

\begin{figure}[t!]
\center
\psfrag{ }[ct][ct][1.]{ }
\includegraphics[width=8cm]{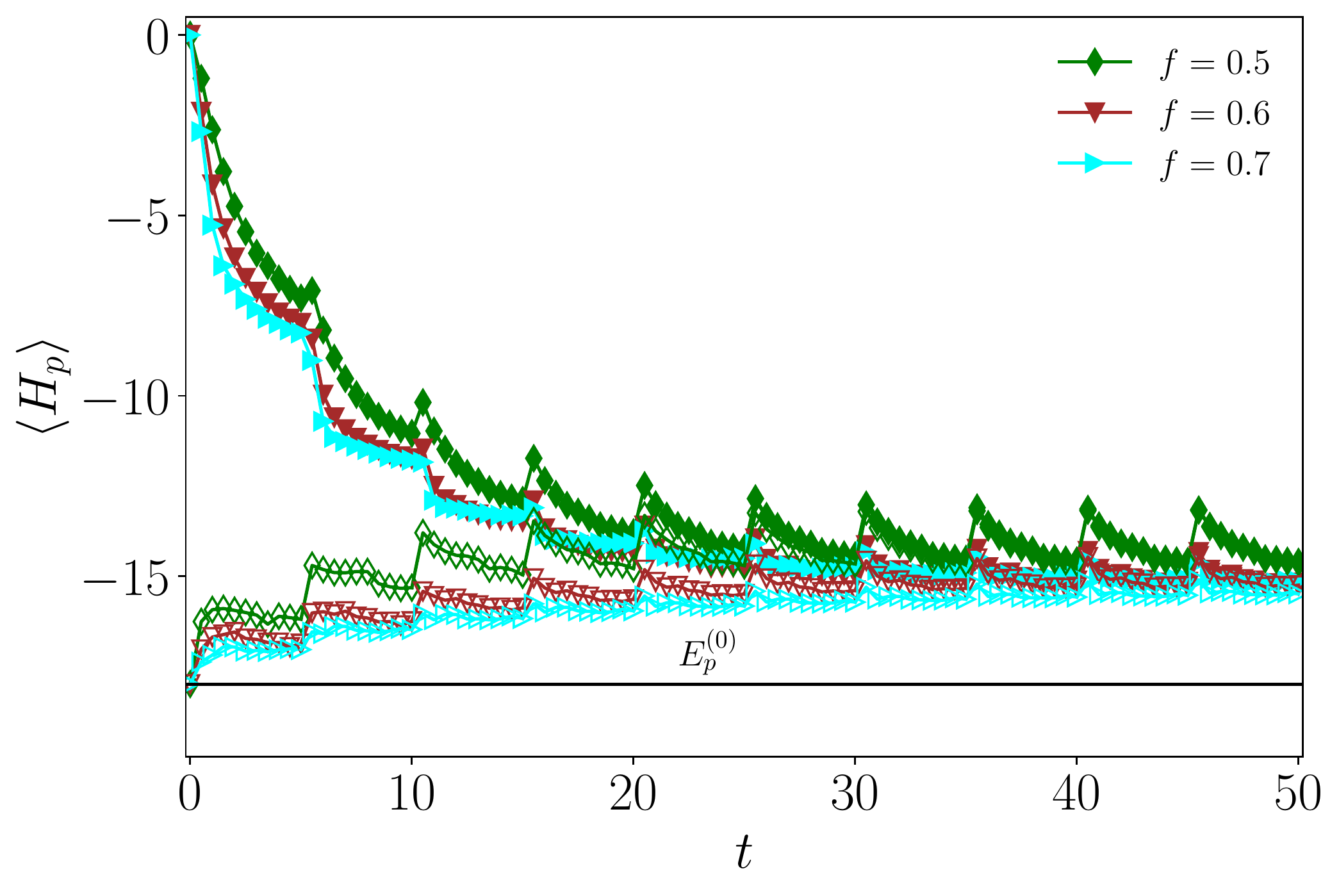}
\includegraphics[width=8cm]{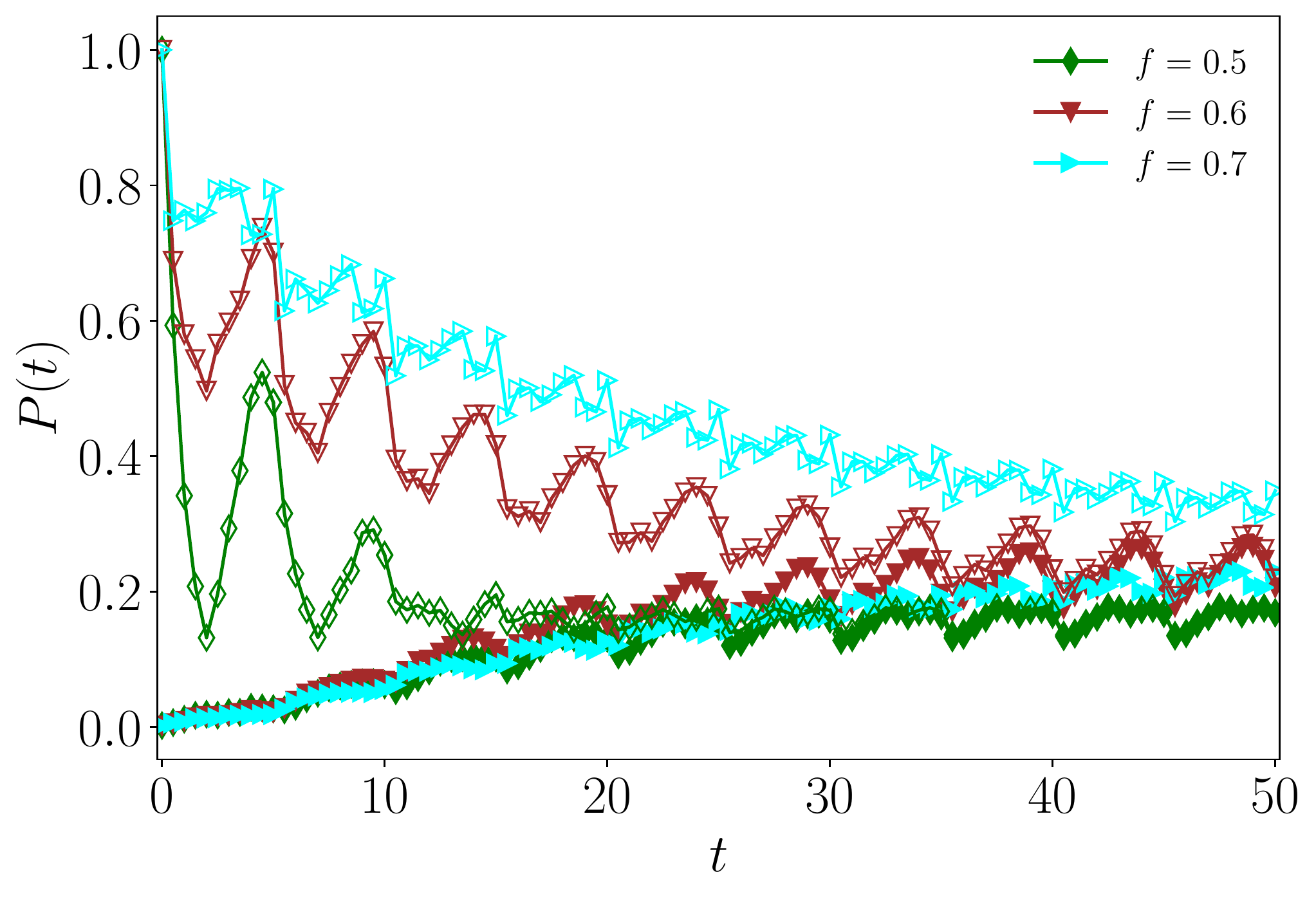}
\caption{Expectation value of the problem Hamiltonian (top) and fidelity (bottom) as a function of $t$ for the cooling protocol consisting of repeated collisions with the auxiliary system \eqref{HA:def} for different values of $f$,  with $\alpha=3$,  $n_c=10$ cooling strokes of duration $\Delta t=5$ in dimensionless time units. The full symbols correspond to the case where the system is initially prepared in the ground state of \eqref{Hd:def}, as in figs.~\ref{fig:rep:coll}--\ref{fig:reduceJ}, while the empty symbols correspond to the case where the system is initially prepared in its own ground state. }
\label{fig:GS}
\end{figure}

In order to check that the repeated collision process works regardless of the initial state, in  Fig.~\ref{fig:GS} we study the case in which the system to be cooled is initially prepared in its own ground state. From the figure, we can see that, after a few cycles, the results for this case and for the case analysed in Fig.~\ref{fig:rep:comp} approach each other and reach approximately the same behaviour within the timescales depicted in the figure.  Only for the large value $f=0.7$, we observe some deviation possibly due to finite sizes. This strongly suggests that the system is reaching some kind of thermal equilibrium, independent of its starting state. This behaviour is fundamentally different from unitary algorithms such as the adiabatic algorithm, where by construction memory of the starting state is maintained throughout the algorithm. It also suggests that the values here are the ultimate limit which the algorithm can achieve and that waiting for longer would not lead to improved performance.

\begin{figure}
\center
\psfrag{ }[ct][ct][1.]{ }
\includegraphics[width=8cm]{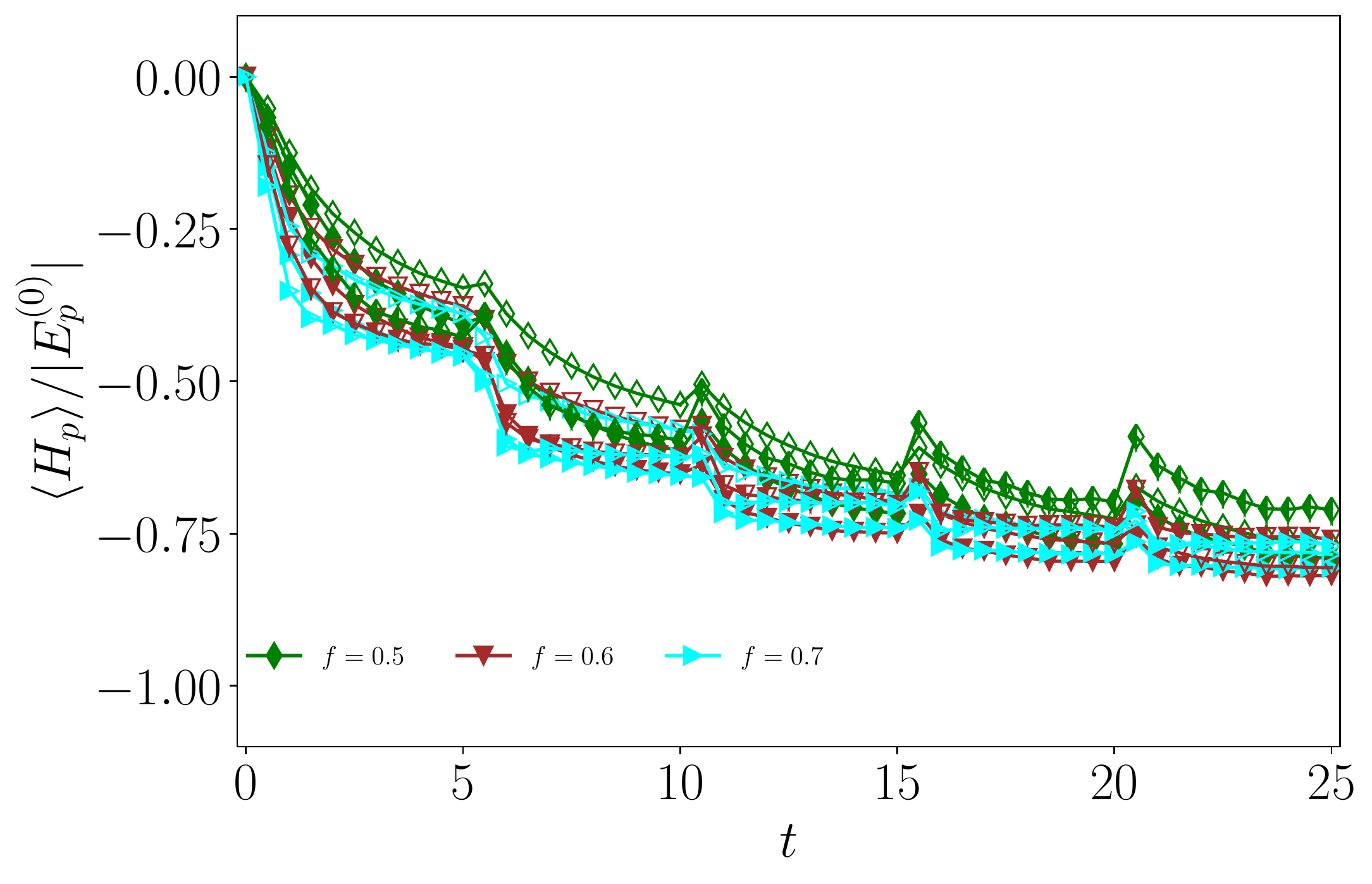}
\caption{Expectation value of the problem Hamiltonian divided by the absolute value of the ground state energy as a function of $t$ for the cooling protocol consisting of repeated collisions with the auxiliary system \eqref{HA:def} for different values of $f$,  with $\alpha=3$,  $n_c=5$ cooling strokes of duration $\Delta t=5$ in dimensionless time units. The different curves with the same color correspond to different realization of the SK model, i.e. to different sets of  $J_{ij}$ and $h_i$, sampled from the same normal distributions. The full symbols correspond to the SK model used in the main text, the empty symbols correspond to two different realizations of the quenched disorder. 
}
\label{fig:seed}
\end{figure}

\alb{In Fig.~\ref{fig:seed} we show the result for the cooling protocol for three different realization of the SK chain, i.e. for different sets of values of $J_{ij}$ and $h_i$ which are drawn from a normal distribution with zero mean and unitary variance. One of the three  sets is the one used in the main text.}

\begin{figure}[h]
\center
\psfrag{ }[ct][ct][1.]{ }
\includegraphics[width=8cm]{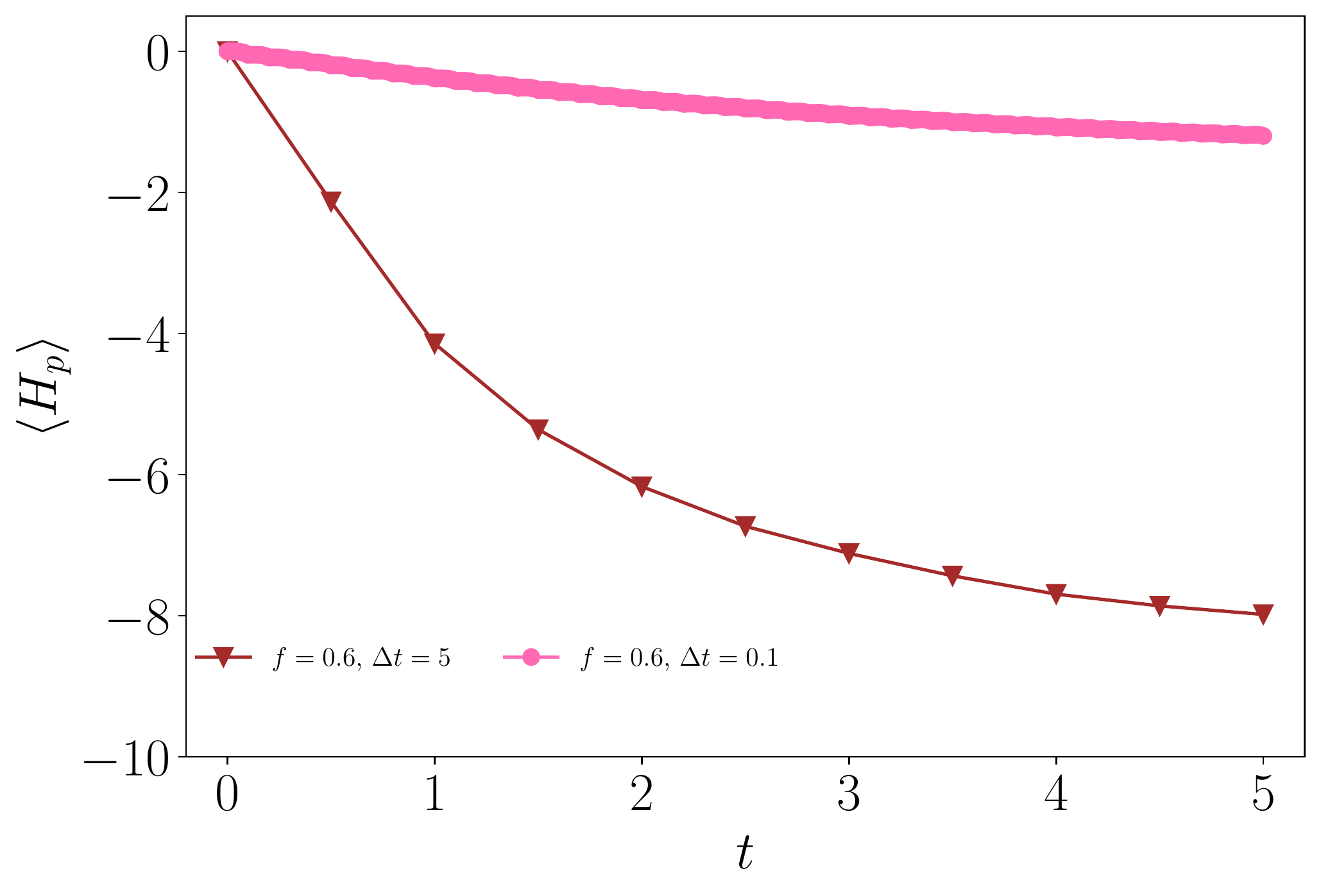}
\caption{Expectation value of the problem Hamiltonian as a function of $t$ for the cooling protocol consisting of collisions with the auxiliary system \eqref{HA:def}. Triangles: single collision of duration $\Delta t=5$ in dimensionless time units,  with $\alpha=3$ as in the main text. Circles: 50 collisions of duration $\Delta t=0.1$,  with $\alpha=3/\sqrt{\Delta t}$. }
\label{short_vs_long}
\end{figure}

\albb{In Fig.~\ref{short_vs_long} we show the difference in cooling performance of the protocol with long ($\Delta t=5$)  and short ($\Delta t=0.1$) bath-system interaction strokes. This elucidates the need of non-Markovian memory effects in the environment. For the long-stroke protocol the prefactor of the interaction Hamiltonian \eqref{HI:def} takes the value $\alpha=3$ as in the main text. In this regime, the environment interacts for a sufficiently long time such that back-flow of information from the environment to the system, a hallmark of non Markovianity, is possible. For the short-stroke protocol, the prefactor of the interaction Hamiltonian  reads $\alpha=3/\sqrt{\Delta t}$. In this regime the protocol dynamics for the system becomes equivalent to the one described by a Markovian GKSL master equation \cite{De_Chiara_2018}. We see that the long-stroke protocol performs considerably better than the short one, indicating that bath and the system need to interact for a finite time for the bath to extract an appreciable amount of heat from the system.}

\bibliography{bibliography}

\end{document}